%% file: main.tex
\documentclass{llncs}

\usepackage[FIGBOTCAP,TABBOTCAP,center,nooneline]{subfigure}
\usepackage[dvips]{rotating}
\usepackage{RSalgorithmic}
\usepackage{algorithm}
\usepackage{amscd}
\usepackage{amsfonts}
\usepackage{amsmath}
\usepackage{amssymb}
\usepackage{booktabs}
\usepackage{color}
\usepackage{datetime}
\usepackage{epsfig}
\usepackage{graphicx}
\usepackage{latexsym}
\usepackage{marginnote}
\usepackage{multirow}
\usepackage{paralist}
\usepackage{placeins}
\usepackage{rotate} 
\usepackage{tikz}
\usepackage{times} 
\usepackage{url}
\usepackage{varwidth}
\usepackage{verbatim} 
\usepackage{wrapfig}
\usepackage{xcolor,colortbl}
\usepackage{xspace}
\usepackage[gen]{eurosym}

\usepackage{tikz}
\usetikzlibrary{arrows}

\input{rs_macros_colors}
\input{rs_macros_dl}
\input{rs_macros_general}


\input{rs_macros_optsmt}


\input{rs_macros_smt}

\input{rs_macros_specific}

\input{mc_macros}

\newcommand{\longversion}{true}
\newcommand{\ignoreinfinal}[1]{}
\ifthenelse{\equal{\longversion}{true}}
{%
  \renewcommand{\ignoreinshort}[1]{\textcolor{black}{#1}}
  \renewcommand{\ignoreinfinal}[1]{\textcolor{black}{#1}}
  \renewcommand{\ignoreinlong}[1]{}
}%
{%
 \renewcommand{\ignoreinfinal}[1]{}
 \renewcommand{\ignoreinshort}[1]{}
 \renewcommand{\ignoreinlong}[1]{\textcolor{black}{#1}}
}

\renewcommand{\TODO}[1]{{\bf \textcolor{blue}{{\fbox{TODO:} #1}}}}

\newcommand{\JMSUGGESTION}[1]{{\bf \textcolor{blue}{{\fbox{JM Directions:} #1}}}}
\renewcommand{\JMSUGGESTION}[1]{}

\newcommand{\FUTUREWORK}[1]{}

\begin{document}

\pagestyle{plain}
\pagenumbering{roman}

\pagestyle{plain}
\pagenumbering{arabic}

\title{%
Requirements Evolution and Evolution Requirements
 \\
 with Constrained Goal Models
\thanks{This research was partially supported by the ERC advanced grant
267856, `Lucretius: Foundations for Software Evolution' and by 
SRC 
GRC Research Project 2012-TJ-2266 WOLF.
}
}

\author{%
Chi Mai Nguyen \and
Roberto Sebastiani \and
Paolo Giorgini \and
John Mylopoulos 
}

\institute{%
DISI, University of Trento, Italy%
}

\maketitle
 \begin{abstract}
 \input{abstract}
 \end{abstract}
 
\renewcommand{\RSTODO}[1]{\noindent{\bf \textcolor{blue}{{\fbox{RS TODO:} #1}}}}


\section{Introduction}
\label{sec:intro}
\input{intro}

\section{Background: Constrained Goal Models}
\label{sec:background}
\input{background}

\input{background_ex}


\section{Requirements Evolution and Evolution Requirements}
\label{sec:evolving}
\input{evolving}

\ignoreinshort{\subsection{Requirements Evolution}}
\label{sec:reqevolution}
\input{reqevolution}

\ignoreinshort{\subsection{Evolution Requirements}}
\label{sec:evolutionreq}
\input{evolutionreq}

\ignoreinshort{\input{appendix}}

\ignoreinfinal{
\section{Automated Reasoning with Evolution Requirements}
\label{sec:evolutionreq_reasoning}
\label{sec:background_formal}
\input{background_formal}
\input{evolutionreq_reasoning}

}
 

\ignoreinfinal{
\section{Implementation}
\label{sec:implementation}
\input{implementation}

}


\ignoreinfinal{
\section{Related Work}
\label{sec:related}
\input{related}
}
\ignoreinfinal{
\section{Conclusions}
\label{sec:concl}
\input{concl}
}

\FloatBarrier
\bibliographystyle{abbrv} 
\bibliography{rs_refs,rs_ownrefs,rs_specific,mc_refs,pg,jm,mc,sathandbook}

\end{document}

%% file: rs_macros_general.tex
\input{rgb}

\newcommand{\new}[1]{{\blue #1}\/}

\newcommand{\resp}[1]{[resp. #1]}

\newcommand{\TODO}[1]{{}}

\newcommand{\ignore}[1]{}
\newcommand{\todo}[1] {}

\newcommand{\RSTODO}[1]{{\bf \textcolor{darkgreen}{{\fbox{RS TODO:} #1}}}}
\renewcommand{\RSTODO}[1]{}

 \newcommand{\ignoreinshort}[1]{}
 \newcommand{\ignoreinlong}[1]{{#1}}

\def\makenewenumerate#1#2{%
\newcounter{cnt#1}
\newenvironment{#1}%
{\begin{list}{\makebox[0pt][r]{#2}}%
{\setlength{\itemsep}{0pt}%
 \setlength{\parsep}{.1em}%
 \setlength{\leftmargin}{1.5em}%
 \setlength{\labelwidth}{.4em}%
 \usecounter{cnt#1}}}
{\end{list}}}

\makenewenumerate{myenumerate}{\arabic{cntmyenumerate}.}

\makenewenumerate{renumerate}{\rm(\roman{cntrenumerate})}
\makenewenumerate{renumerateprime}{\rm(\roman{cntrenumerateprime}$'$)}
\makenewenumerate{renumeratesecond}{\rm(\roman{cntrenumeratesecond}$''$)}

\makenewenumerate{aenumerate}{({\it\alph{cntaenumerate}})}
\makenewenumerate{aenumerateprime}{({\it\alph{cntaenumerateprime}$'$})}
\makenewenumerate{aenumeratesecond}{({\it\alph{cntaenumeratesecond}$''$})}

\def\newplaintheorem#1#2{%
\newtheorem{#1plain}{#2}[section]%
\newenvironment{#1}{\begin{#1plain}\rm }{\end{#1plain}}}

\newplaintheorem{notation}{Notation}
\newplaintheorem{cexample}{Counter-Example}

\newcommand{\fakesubsubsection}[1]{\smallskip\noindent {\bf #1.}}

\newcommand{\sref}[1]{\S{}\ref{#1}}
\newcommand{\noi}{\noindent}

\newcommand{\tuple}[1]{\ensuremath{\langle{#1}\rangle}\xspace}
\newcommand{\set}[1]{\ensuremath{\{{#1}\}}\xspace}
\newcommand{\imp}{\ensuremath{\rightarrow}\xspace}

\renewcommand{\iff}{\ensuremath{\leftrightarrow}\xspace}
\newcommand{\defas}{\ensuremath{\stackrel{\text{\tiny def}}{=}}\xspace}

\newcommand\cala{\ensuremath{\mathcal{A}}\xspace}
\newcommand\calb{\ensuremath{\mathcal{B}}\xspace}

\newcommand\cald{\ensuremath{\mathcal{D}}\xspace}
\newcommand\cale{\ensuremath{\mathcal{E}}\xspace}

\newcommand\calg{\ensuremath{\mathcal{G}}\xspace}

\newcommand\calm{\ensuremath{\mathcal{M}}\xspace}
\newcommand\caln{\ensuremath{\mathcal{N}}\xspace}

\newcommand\calr{\ensuremath{\mathcal{R}}\xspace}



%% file: rgb.tex
\definecolor {snow}                {rgb}{1.00,0.98,0.98}
\definecolor {ghostwhite}          {rgb}{0.97,0.97,1.00}
\definecolor {whitesmoke}          {rgb}{0.96,0.96,0.96}
\definecolor {gainsboro}           {rgb}{0.86,0.86,0.86}
\definecolor {floralwhite}         {rgb}{1.00,0.98,0.94}
\definecolor {oldlace}             {rgb}{0.99,0.96,0.90}
\definecolor {linen}               {rgb}{0.98,0.94,0.90}
\definecolor {antiquewhite}        {rgb}{0.98,0.92,0.84}
\definecolor {papayawhip}          {rgb}{1.00,0.94,0.84}
\definecolor {blanchedalmond}      {rgb}{1.00,0.92,0.80}
\definecolor {bisque}              {rgb}{1.00,0.89,0.77}
\definecolor {peachpuff}           {rgb}{1.00,0.85,0.73}
\definecolor {navajowhite}         {rgb}{1.00,0.87,0.68}
\definecolor {moccasin}            {rgb}{1.00,0.89,0.71}
\definecolor {cornsilk}            {rgb}{1.00,0.97,0.86}
\definecolor {ivory}               {rgb}{1.00,1.00,0.94}
\definecolor {lemonchiffon}        {rgb}{1.00,0.98,0.80}
\definecolor {seashell}            {rgb}{1.00,0.96,0.93}
\definecolor {honeydew}            {rgb}{0.94,1.00,0.94}
\definecolor {mintcream}           {rgb}{0.96,1.00,0.98}
\definecolor {azure}               {rgb}{0.94,1.00,1.00}
\definecolor {aliceblue}           {rgb}{0.94,0.97,1.00}
\definecolor {lavender}            {rgb}{0.90,0.90,0.98}
\definecolor {lavenderblush}       {rgb}{1.00,0.94,0.96}
\definecolor {mistyrose}           {rgb}{1.00,0.89,0.88}
\definecolor {white}               {rgb}{1.00,1.00,1.00}
\definecolor {black}               {rgb}{0.00,0.00,0.00}
\definecolor {darkslategray}       {rgb}{0.18,0.31,0.31}
\definecolor {dimgray}             {rgb}{0.41,0.41,0.41}
\definecolor {slategray}           {rgb}{0.44,0.50,0.56}
\definecolor {lightslategray}      {rgb}{0.47,0.53,0.60}
\definecolor {gray}                {rgb}{0.75,0.75,0.75}
\definecolor {lightgrey}           {rgb}{0.83,0.83,0.83}
\definecolor {midnightblue}        {rgb}{0.10,0.10,0.44}
\definecolor {navy}                {rgb}{0.00,0.00,0.50}
\definecolor {cornflowerblue}      {rgb}{0.39,0.58,0.93}
\definecolor {darkslateblue}       {rgb}{0.28,0.24,0.55}
\definecolor {slateblue}           {rgb}{0.42,0.35,0.80}
\definecolor {mediumslateblue}     {rgb}{0.48,0.41,0.93}
\definecolor {lightslateblue}      {rgb}{0.52,0.44,1.00}
\definecolor {mediumblue}          {rgb}{0.00,0.00,0.80}
\definecolor {royalblue}           {rgb}{0.25,0.41,0.88}
\definecolor {blue}                {rgb}{0.00,0.00,1.00}
\definecolor {dodgerblue}          {rgb}{0.12,0.56,1.00}
\definecolor {deepskyblue}         {rgb}{0.00,0.75,1.00}
\definecolor {skyblue}             {rgb}{0.53,0.81,0.92}
\definecolor {lightskyblue}        {rgb}{0.53,0.81,0.98}
\definecolor {steelblue}           {rgb}{0.27,0.51,0.71}
\definecolor {lightsteelblue}      {rgb}{0.69,0.77,0.87}
\definecolor {lightblue}           {rgb}{0.68,0.85,0.90}
\definecolor {powderblue}          {rgb}{0.69,0.88,0.90}
\definecolor {paleturquoise}       {rgb}{0.69,0.93,0.93}
\definecolor {darkturquoise}       {rgb}{0.00,0.81,0.82}
\definecolor {mediumturquoise}     {rgb}{0.28,0.82,0.80}
\definecolor {turquoise}           {rgb}{0.25,0.88,0.82}
\definecolor {cyan}                {rgb}{0.00,1.00,1.00}
\definecolor {lightcyan}           {rgb}{0.88,1.00,1.00}
\definecolor {cadetblue}           {rgb}{0.37,0.62,0.63}
\definecolor {mediumaquamarine}    {rgb}{0.40,0.80,0.67}
\definecolor {aquamarine}          {rgb}{0.50,1.00,0.83}
\definecolor {darkgreen}           {rgb}{0.00,0.39,0.00}
\definecolor {darkolivegreen}      {rgb}{0.33,0.42,0.18}
\definecolor {darkseagreen}        {rgb}{0.56,0.74,0.56}
\definecolor {seagreen}            {rgb}{0.18,0.55,0.34}
\definecolor {mediumseagreen}      {rgb}{0.24,0.70,0.44}
\definecolor {lightseagreen}       {rgb}{0.13,0.70,0.67}
\definecolor {palegreen}           {rgb}{0.60,0.98,0.60}
\definecolor {springgreen}         {rgb}{0.00,1.00,0.50}
\definecolor {lawngreen}           {rgb}{0.49,0.99,0.00}
\definecolor {green}               {rgb}{0.00,1.00,0.00}
\definecolor {chartreuse}          {rgb}{0.50,1.00,0.00}
\definecolor {mediumspringgreen}   {rgb}{0.00,0.98,0.60}
\definecolor {greenyellow}         {rgb}{0.68,1.00,0.18}
\definecolor {limegreen}           {rgb}{0.20,0.80,0.20}
\definecolor {yellowgreen}         {rgb}{0.60,0.80,0.20}
\definecolor {forestgreen}         {rgb}{0.13,0.55,0.13}
\definecolor {olivedrab}           {rgb}{0.42,0.56,0.14}
\definecolor {darkkhaki}           {rgb}{0.74,0.72,0.42}
\definecolor {khaki}               {rgb}{0.94,0.90,0.55}
\definecolor {palegoldenrod}       {rgb}{0.93,0.91,0.67}
\definecolor {lightgoldenrodyellow} {rgb}{0.98,0.98,0.82}
\definecolor {lightyellow}         {rgb}{1.00,1.00,0.88}
\definecolor {yellow}              {rgb}{1.00,1.00,0.00}
\definecolor {gold}                {rgb}{1.00,0.84,0.00}
\definecolor {lightgoldenrod}      {rgb}{0.93,0.87,0.51}
\definecolor {goldenrod}           {rgb}{0.85,0.65,0.13}
\definecolor {darkgoldenrod}       {rgb}{0.72,0.53,0.04}
\definecolor {rosybrown}           {rgb}{0.74,0.56,0.56}
\definecolor {indianred}           {rgb}{0.80,0.36,0.36}
\definecolor {saddlebrown}         {rgb}{0.55,0.27,0.07}
\definecolor {sienna}              {rgb}{0.63,0.32,0.18}
\definecolor {peru}                {rgb}{0.80,0.52,0.25}
\definecolor {burlywood}           {rgb}{0.87,0.72,0.53}
\definecolor {beige}               {rgb}{0.96,0.96,0.86}
\definecolor {wheat}               {rgb}{0.96,0.87,0.70}
\definecolor {sandybrown}          {rgb}{0.96,0.64,0.38}
\definecolor {tan}                 {rgb}{0.82,0.71,0.55}
\definecolor {chocolate}           {rgb}{0.82,0.41,0.12}
\definecolor {firebrick}           {rgb}{0.70,0.13,0.13}
\definecolor {brown}               {rgb}{0.65,0.16,0.16}
\definecolor {darksalmon}          {rgb}{0.91,0.59,0.48}
\definecolor {salmon}              {rgb}{0.98,0.50,0.45}
\definecolor {lightsalmon}         {rgb}{1.00,0.63,0.48}
\definecolor {orange}              {rgb}{1.00,0.65,0.00}
\definecolor {darkorange}          {rgb}{1.00,0.55,0.00}
\definecolor {coral}               {rgb}{1.00,0.50,0.31}
\definecolor {lightcoral}          {rgb}{0.94,0.50,0.50}
\definecolor {tomato}              {rgb}{1.00,0.39,0.28}
\definecolor {orangered}           {rgb}{1.00,0.27,0.00}
\definecolor {red}                 {rgb}{1.00,0.00,0.00}
\definecolor {hotpink}             {rgb}{1.00,0.41,0.71}
\definecolor {deeppink}            {rgb}{1.00,0.08,0.58}
\definecolor {pink}                {rgb}{1.00,0.75,0.80}
\definecolor {lightpink}           {rgb}{1.00,0.71,0.76}
\definecolor {palevioletred}       {rgb}{0.86,0.44,0.58}
\definecolor {maroon}              {rgb}{0.69,0.19,0.38}
\definecolor {mediumvioletred}     {rgb}{0.78,0.08,0.52}
\definecolor {violetred}           {rgb}{0.82,0.13,0.56}
\definecolor {magenta}             {rgb}{1.00,0.00,1.00}
\definecolor {violet}              {rgb}{0.93,0.51,0.93}
\definecolor {plum}                {rgb}{0.87,0.63,0.87}
\definecolor {orchid}              {rgb}{0.85,0.44,0.84}
\definecolor {mediumorchid}        {rgb}{0.73,0.33,0.83}
\definecolor {darkorchid}          {rgb}{0.60,0.20,0.80}
\definecolor {darkviolet}          {rgb}{0.58,0.00,0.83}
\definecolor {blueviolet}          {rgb}{0.54,0.17,0.89}
\definecolor {purple}              {rgb}{0.63,0.13,0.94}
\definecolor {mediumpurple}        {rgb}{0.58,0.44,0.86}
\definecolor {thistle}             {rgb}{0.85,0.75,0.85}
\definecolor {snow2}               {rgb}{0.93,0.91,0.91}
\definecolor {snow3}               {rgb}{0.80,0.79,0.79}
\definecolor {snow4}               {rgb}{0.55,0.54,0.54}
\definecolor {seashell2}           {rgb}{0.93,0.90,0.87}
\definecolor {seashell3}           {rgb}{0.80,0.77,0.75}
\definecolor {seashell4}           {rgb}{0.55,0.53,0.51}
\definecolor {antiquewhite1}       {rgb}{1.00,0.94,0.86}
\definecolor {antiquewhite2}       {rgb}{0.93,0.87,0.80}
\definecolor {antiquewhite3}       {rgb}{0.80,0.75,0.69}
\definecolor {antiquewhite4}       {rgb}{0.55,0.51,0.47}
\definecolor {bisque2}             {rgb}{0.93,0.84,0.72}
\definecolor {bisque3}             {rgb}{0.80,0.72,0.62}
\definecolor {bisque4}             {rgb}{0.55,0.49,0.42}
\definecolor {peachpuff2}          {rgb}{0.93,0.80,0.68}
\definecolor {peachpuff3}          {rgb}{0.80,0.69,0.58}
\definecolor {peachpuff4}          {rgb}{0.55,0.47,0.40}
\definecolor {navajowhite2}        {rgb}{0.93,0.81,0.63}
\definecolor {navajowhite3}        {rgb}{0.80,0.70,0.55}
\definecolor {navajowhite4}        {rgb}{0.55,0.47,0.37}
\definecolor {lemonchiffon2}       {rgb}{0.93,0.91,0.75}
\definecolor {lemonchiffon3}       {rgb}{0.80,0.79,0.65}
\definecolor {lemonchiffon4}       {rgb}{0.55,0.54,0.44}
\definecolor {cornsilk2}           {rgb}{0.93,0.91,0.80}
\definecolor {cornsilk3}           {rgb}{0.80,0.78,0.69}
\definecolor {cornsilk4}           {rgb}{0.55,0.53,0.47}
\definecolor {ivory2}              {rgb}{0.93,0.93,0.88}
\definecolor {ivory3}              {rgb}{0.80,0.80,0.76}
\definecolor {ivory4}              {rgb}{0.55,0.55,0.51}
\definecolor {honeydew2}           {rgb}{0.88,0.93,0.88}
\definecolor {honeydew3}           {rgb}{0.76,0.80,0.76}
\definecolor {honeydew4}           {rgb}{0.51,0.55,0.51}
\definecolor {lavenderblush2}      {rgb}{0.93,0.88,0.90}
\definecolor {lavenderblush3}      {rgb}{0.80,0.76,0.77}
\definecolor {lavenderblush4}      {rgb}{0.55,0.51,0.53}
\definecolor {mistyrose2}          {rgb}{0.93,0.84,0.82}
\definecolor {mistyrose3}          {rgb}{0.80,0.72,0.71}
\definecolor {mistyrose4}          {rgb}{0.55,0.49,0.48}
\definecolor {azure2}              {rgb}{0.88,0.93,0.93}
\definecolor {azure3}              {rgb}{0.76,0.80,0.80}
\definecolor {azure4}              {rgb}{0.51,0.55,0.55}
\definecolor {slateblue1}          {rgb}{0.51,0.44,1.00}
\definecolor {slateblue2}          {rgb}{0.48,0.40,0.93}
\definecolor {slateblue3}          {rgb}{0.41,0.35,0.80}
\definecolor {slateblue4}          {rgb}{0.28,0.24,0.55}
\definecolor {royalblue1}          {rgb}{0.28,0.46,1.00}
\definecolor {royalblue2}          {rgb}{0.26,0.43,0.93}
\definecolor {royalblue3}          {rgb}{0.23,0.37,0.80}
\definecolor {royalblue4}          {rgb}{0.15,0.25,0.55}
\definecolor {blue2}               {rgb}{0.00,0.00,0.93}
\definecolor {blue4}               {rgb}{0.00,0.00,0.55}
\definecolor {dodgerblue2}         {rgb}{0.11,0.53,0.93}
\definecolor {dodgerblue3}         {rgb}{0.09,0.45,0.80}
\definecolor {dodgerblue4}         {rgb}{0.06,0.31,0.55}
\definecolor {steelblue1}          {rgb}{0.39,0.72,1.00}
\definecolor {steelblue2}          {rgb}{0.36,0.67,0.93}
\definecolor {steelblue3}          {rgb}{0.31,0.58,0.80}
\definecolor {steelblue4}          {rgb}{0.21,0.39,0.55}
\definecolor {deepskyblue2}        {rgb}{0.00,0.70,0.93}
\definecolor {deepskyblue3}        {rgb}{0.00,0.60,0.80}
\definecolor {deepskyblue4}        {rgb}{0.00,0.41,0.55}
\definecolor {skyblue1}            {rgb}{0.53,0.81,1.00}
\definecolor {skyblue2}            {rgb}{0.49,0.75,0.93}
\definecolor {skyblue3}            {rgb}{0.42,0.65,0.80}
\definecolor {skyblue4}            {rgb}{0.29,0.44,0.55}
\definecolor {lightskyblue1}       {rgb}{0.69,0.89,1.00}
\definecolor {lightskyblue2}       {rgb}{0.64,0.83,0.93}
\definecolor {lightskyblue3}       {rgb}{0.55,0.71,0.80}
\definecolor {lightskyblue4}       {rgb}{0.38,0.48,0.55}
\definecolor {slategray1}          {rgb}{0.78,0.89,1.00}
\definecolor {slategray2}          {rgb}{0.73,0.83,0.93}
\definecolor {slategray3}          {rgb}{0.62,0.71,0.80}
\definecolor {slategray4}          {rgb}{0.42,0.48,0.55}
\definecolor {lightsteelblue1}     {rgb}{0.79,0.88,1.00}
\definecolor {lightsteelblue2}     {rgb}{0.74,0.82,0.93}
\definecolor {lightsteelblue3}     {rgb}{0.64,0.71,0.80}
\definecolor {lightsteelblue4}     {rgb}{0.43,0.48,0.55}
\definecolor {lightblue1}          {rgb}{0.75,0.94,1.00}
\definecolor {lightblue2}          {rgb}{0.70,0.87,0.93}
\definecolor {lightblue3}          {rgb}{0.60,0.75,0.80}
\definecolor {lightblue4}          {rgb}{0.41,0.51,0.55}
\definecolor {lightcyan2}          {rgb}{0.82,0.93,0.93}
\definecolor {lightcyan3}          {rgb}{0.71,0.80,0.80}
\definecolor {lightcyan4}          {rgb}{0.48,0.55,0.55}
\definecolor {paleturquoise1}      {rgb}{0.73,1.00,1.00}
\definecolor {paleturquoise2}      {rgb}{0.68,0.93,0.93}
\definecolor {paleturquoise3}      {rgb}{0.59,0.80,0.80}
\definecolor {paleturquoise4}      {rgb}{0.40,0.55,0.55}
\definecolor {cadetblue1}          {rgb}{0.60,0.96,1.00}
\definecolor {cadetblue2}          {rgb}{0.56,0.90,0.93}
\definecolor {cadetblue3}          {rgb}{0.48,0.77,0.80}
\definecolor {cadetblue4}          {rgb}{0.33,0.53,0.55}
\definecolor {turquoise1}          {rgb}{0.00,0.96,1.00}
\definecolor {turquoise2}          {rgb}{0.00,0.90,0.93}
\definecolor {turquoise3}          {rgb}{0.00,0.77,0.80}
\definecolor {turquoise4}          {rgb}{0.00,0.53,0.55}
\definecolor {cyan2}               {rgb}{0.00,0.93,0.93}
\definecolor {cyan3}               {rgb}{0.00,0.80,0.80}
\definecolor {cyan4}               {rgb}{0.00,0.55,0.55}
\definecolor {darkslategray1}      {rgb}{0.59,1.00,1.00}
\definecolor {darkslategray2}      {rgb}{0.55,0.93,0.93}
\definecolor {darkslategray3}      {rgb}{0.47,0.80,0.80}
\definecolor {darkslategray4}      {rgb}{0.32,0.55,0.55}
\definecolor {aquamarine2}         {rgb}{0.46,0.93,0.78}
\definecolor {aquamarine4}         {rgb}{0.27,0.55,0.45}
\definecolor {darkseagreen1}       {rgb}{0.76,1.00,0.76}
\definecolor {darkseagreen2}       {rgb}{0.71,0.93,0.71}
\definecolor {darkseagreen3}       {rgb}{0.61,0.80,0.61}
\definecolor {darkseagreen4}       {rgb}{0.41,0.55,0.41}
\definecolor {seagreen1}           {rgb}{0.33,1.00,0.62}
\definecolor {seagreen2}           {rgb}{0.31,0.93,0.58}
\definecolor {seagreen3}           {rgb}{0.26,0.80,0.50}
\definecolor {palegreen1}          {rgb}{0.60,1.00,0.60}
\definecolor {palegreen2}          {rgb}{0.56,0.93,0.56}
\definecolor {palegreen3}          {rgb}{0.49,0.80,0.49}
\definecolor {palegreen4}          {rgb}{0.33,0.55,0.33}
\definecolor {springgreen2}        {rgb}{0.00,0.93,0.46}
\definecolor {springgreen3}        {rgb}{0.00,0.80,0.40}
\definecolor {springgreen4}        {rgb}{0.00,0.55,0.27}
\definecolor {green2}              {rgb}{0.00,0.93,0.00}
\definecolor {green3}              {rgb}{0.00,0.80,0.00}
\definecolor {green4}              {rgb}{0.00,0.55,0.00}
\definecolor {chartreuse2}         {rgb}{0.46,0.93,0.00}
\definecolor {chartreuse3}         {rgb}{0.40,0.80,0.00}
\definecolor {chartreuse4}         {rgb}{0.27,0.55,0.00}
\definecolor {olivedrab1}          {rgb}{0.75,1.00,0.24}
\definecolor {olivedrab2}          {rgb}{0.70,0.93,0.23}
\definecolor {olivedrab4}          {rgb}{0.41,0.55,0.13}
\definecolor {darkolivegreen1}     {rgb}{0.79,1.00,0.44}
\definecolor {darkolivegreen2}     {rgb}{0.74,0.93,0.41}
\definecolor {darkolivegreen3}     {rgb}{0.64,0.80,0.35}
\definecolor {darkolivegreen4}     {rgb}{0.43,0.55,0.24}
\definecolor {khaki1}              {rgb}{1.00,0.96,0.56}
\definecolor {khaki2}              {rgb}{0.93,0.90,0.52}
\definecolor {khaki3}              {rgb}{0.80,0.78,0.45}
\definecolor {khaki4}              {rgb}{0.55,0.53,0.31}
\definecolor {lightgoldenrod1}     {rgb}{1.00,0.93,0.55}
\definecolor {lightgoldenrod2}     {rgb}{0.93,0.86,0.51}
\definecolor {lightgoldenrod3}     {rgb}{0.80,0.75,0.44}
\definecolor {lightgoldenrod4}     {rgb}{0.55,0.51,0.30}
\definecolor {lightyellow2}        {rgb}{0.93,0.93,0.82}
\definecolor {lightyellow3}        {rgb}{0.80,0.80,0.71}
\definecolor {lightyellow4}        {rgb}{0.55,0.55,0.48}
\definecolor {yellow2}             {rgb}{0.93,0.93,0.00}
\definecolor {yellow3}             {rgb}{0.80,0.80,0.00}
\definecolor {yellow4}             {rgb}{0.55,0.55,0.00}
\definecolor {gold2}               {rgb}{0.93,0.79,0.00}
\definecolor {gold3}               {rgb}{0.80,0.68,0.00}
\definecolor {gold4}               {rgb}{0.55,0.46,0.00}
\definecolor {goldenrod1}          {rgb}{1.00,0.76,0.15}
\definecolor {goldenrod2}          {rgb}{0.93,0.71,0.13}
\definecolor {goldenrod3}          {rgb}{0.80,0.61,0.11}
\definecolor {goldenrod4}          {rgb}{0.55,0.41,0.08}
\definecolor {darkgoldenrod1}      {rgb}{1.00,0.73,0.06}
\definecolor {darkgoldenrod2}      {rgb}{0.93,0.68,0.05}
\definecolor {darkgoldenrod3}      {rgb}{0.80,0.58,0.05}
\definecolor {darkgoldenrod4}      {rgb}{0.55,0.40,0.03}
\definecolor {rosybrown1}          {rgb}{1.00,0.76,0.76}
\definecolor {rosybrown2}          {rgb}{0.93,0.71,0.71}
\definecolor {rosybrown3}          {rgb}{0.80,0.61,0.61}
\definecolor {rosybrown4}          {rgb}{0.55,0.41,0.41}
\definecolor {indianred1}          {rgb}{1.00,0.42,0.42}
\definecolor {indianred2}          {rgb}{0.93,0.39,0.39}
\definecolor {indianred3}          {rgb}{0.80,0.33,0.33}
\definecolor {indianred4}          {rgb}{0.55,0.23,0.23}
\definecolor {sienna1}             {rgb}{1.00,0.51,0.28}
\definecolor {sienna2}             {rgb}{0.93,0.47,0.26}
\definecolor {sienna3}             {rgb}{0.80,0.41,0.22}
\definecolor {sienna4}             {rgb}{0.55,0.28,0.15}
\definecolor {burlywood1}          {rgb}{1.00,0.83,0.61}
\definecolor {burlywood2}          {rgb}{0.93,0.77,0.57}
\definecolor {burlywood3}          {rgb}{0.80,0.67,0.49}
\definecolor {burlywood4}          {rgb}{0.55,0.45,0.33}
\definecolor {wheat1}              {rgb}{1.00,0.91,0.73}
\definecolor {wheat2}              {rgb}{0.93,0.85,0.68}
\definecolor {wheat3}              {rgb}{0.80,0.73,0.59}
\definecolor {wheat4}              {rgb}{0.55,0.49,0.40}
\definecolor {tan1}                {rgb}{1.00,0.65,0.31}
\definecolor {tan2}                {rgb}{0.93,0.60,0.29}
\definecolor {tan4}                {rgb}{0.55,0.35,0.17}
\definecolor {chocolate1}          {rgb}{1.00,0.50,0.14}
\definecolor {chocolate2}          {rgb}{0.93,0.46,0.13}
\definecolor {chocolate3}          {rgb}{0.80,0.40,0.11}
\definecolor {firebrick1}          {rgb}{1.00,0.19,0.19}
\definecolor {firebrick2}          {rgb}{0.93,0.17,0.17}
\definecolor {firebrick3}          {rgb}{0.80,0.15,0.15}
\definecolor {firebrick4}          {rgb}{0.55,0.10,0.10}
\definecolor {brown1}              {rgb}{1.00,0.25,0.25}
\definecolor {brown2}              {rgb}{0.93,0.23,0.23}
\definecolor {brown3}              {rgb}{0.80,0.20,0.20}
\definecolor {brown4}              {rgb}{0.55,0.14,0.14}
\definecolor {salmon1}             {rgb}{1.00,0.55,0.41}
\definecolor {salmon2}             {rgb}{0.93,0.51,0.38}
\definecolor {salmon3}             {rgb}{0.80,0.44,0.33}
\definecolor {salmon4}             {rgb}{0.55,0.30,0.22}
\definecolor {lightsalmon2}        {rgb}{0.93,0.58,0.45}
\definecolor {lightsalmon3}        {rgb}{0.80,0.51,0.38}
\definecolor {lightsalmon4}        {rgb}{0.55,0.34,0.26}
\definecolor {orange2}             {rgb}{0.93,0.60,0.00}
\definecolor {orange3}             {rgb}{0.80,0.52,0.00}
\definecolor {orange4}             {rgb}{0.55,0.35,0.00}
\definecolor {darkorange1}         {rgb}{1.00,0.50,0.00}
\definecolor {darkorange2}         {rgb}{0.93,0.46,0.00}
\definecolor {darkorange3}         {rgb}{0.80,0.40,0.00}
\definecolor {darkorange4}         {rgb}{0.55,0.27,0.00}
\definecolor {coral1}              {rgb}{1.00,0.45,0.34}
\definecolor {coral2}              {rgb}{0.93,0.42,0.31}
\definecolor {coral3}              {rgb}{0.80,0.36,0.27}
\definecolor {coral4}              {rgb}{0.55,0.24,0.18}
\definecolor {tomato2}             {rgb}{0.93,0.36,0.26}
\definecolor {tomato3}             {rgb}{0.80,0.31,0.22}
\definecolor {tomato4}             {rgb}{0.55,0.21,0.15}
\definecolor {orangered2}          {rgb}{0.93,0.25,0.00}
\definecolor {orangered3}          {rgb}{0.80,0.22,0.00}
\definecolor {orangered4}          {rgb}{0.55,0.15,0.00}
\definecolor {red2}                {rgb}{0.93,0.00,0.00}
\definecolor {red3}                {rgb}{0.80,0.00,0.00}
\definecolor {red4}                {rgb}{0.55,0.00,0.00}
\definecolor {deeppink2}           {rgb}{0.93,0.07,0.54}
\definecolor {deeppink3}           {rgb}{0.80,0.06,0.46}
\definecolor {deeppink4}           {rgb}{0.55,0.04,0.31}
\definecolor {hotpink1}            {rgb}{1.00,0.43,0.71}
\definecolor {hotpink2}            {rgb}{0.93,0.42,0.65}
\definecolor {hotpink3}            {rgb}{0.80,0.38,0.56}
\definecolor {hotpink4}            {rgb}{0.55,0.23,0.38}
\definecolor {pink1}               {rgb}{1.00,0.71,0.77}
\definecolor {pink2}               {rgb}{0.93,0.66,0.72}
\definecolor {pink3}               {rgb}{0.80,0.57,0.62}
\definecolor {pink4}               {rgb}{0.55,0.39,0.42}
\definecolor {lightpink1}          {rgb}{1.00,0.68,0.73}
\definecolor {lightpink2}          {rgb}{0.93,0.64,0.68}
\definecolor {lightpink3}          {rgb}{0.80,0.55,0.58}
\definecolor {lightpink4}          {rgb}{0.55,0.37,0.40}
\definecolor {palevioletred1}      {rgb}{1.00,0.51,0.67}
\definecolor {palevioletred2}      {rgb}{0.93,0.47,0.62}
\definecolor {palevioletred3}      {rgb}{0.80,0.41,0.54}
\definecolor {palevioletred4}      {rgb}{0.55,0.28,0.36}
\definecolor {maroon1}             {rgb}{1.00,0.20,0.70}
\definecolor {maroon2}             {rgb}{0.93,0.19,0.65}
\definecolor {maroon3}             {rgb}{0.80,0.16,0.56}
\definecolor {maroon4}             {rgb}{0.55,0.11,0.38}
\definecolor {violetred1}          {rgb}{1.00,0.24,0.59}
\definecolor {violetred2}          {rgb}{0.93,0.23,0.55}
\definecolor {violetred3}          {rgb}{0.80,0.20,0.47}
\definecolor {violetred4}          {rgb}{0.55,0.13,0.32}
\definecolor {magenta2}            {rgb}{0.93,0.00,0.93}
\definecolor {magenta3}            {rgb}{0.80,0.00,0.80}
\definecolor {magenta4}            {rgb}{0.55,0.00,0.55}
\definecolor {orchid1}             {rgb}{1.00,0.51,0.98}
\definecolor {orchid2}             {rgb}{0.93,0.48,0.91}
\definecolor {orchid3}             {rgb}{0.80,0.41,0.79}
\definecolor {orchid4}             {rgb}{0.55,0.28,0.54}
\definecolor {plum1}               {rgb}{1.00,0.73,1.00}
\definecolor {plum2}               {rgb}{0.93,0.68,0.93}
\definecolor {plum3}               {rgb}{0.80,0.59,0.80}
\definecolor {plum4}               {rgb}{0.55,0.40,0.55}
\definecolor {mediumorchid1}       {rgb}{0.88,0.40,1.00}
\definecolor {mediumorchid2}       {rgb}{0.82,0.37,0.93}
\definecolor {mediumorchid3}       {rgb}{0.71,0.32,0.80}
\definecolor {mediumorchid4}       {rgb}{0.48,0.22,0.55}
\definecolor {darkorchid1}         {rgb}{0.75,0.24,1.00}
\definecolor {darkorchid2}         {rgb}{0.70,0.23,0.93}
\definecolor {darkorchid3}         {rgb}{0.60,0.20,0.80}
\definecolor {darkorchid4}         {rgb}{0.41,0.13,0.55}
\definecolor {purple1}             {rgb}{0.61,0.19,1.00}
\definecolor {purple2}             {rgb}{0.57,0.17,0.93}
\definecolor {purple3}             {rgb}{0.49,0.15,0.80}
\definecolor {purple4}             {rgb}{0.33,0.10,0.55}
\definecolor {mediumpurple1}       {rgb}{0.67,0.51,1.00}
\definecolor {mediumpurple2}       {rgb}{0.62,0.47,0.93}
\definecolor {mediumpurple3}       {rgb}{0.54,0.41,0.80}
\definecolor {mediumpurple4}       {rgb}{0.36,0.28,0.55}
\definecolor {thistle1}            {rgb}{1.00,0.88,1.00}
\definecolor {thistle2}            {rgb}{0.93,0.82,0.93}
\definecolor {thistle3}            {rgb}{0.80,0.71,0.80}
\definecolor {thistle4}            {rgb}{0.55,0.48,0.55}
\definecolor {gray1}               {rgb}{0.01,0.01,0.01}
\definecolor {gray2}               {rgb}{0.02,0.02,0.02}
\definecolor {gray3}               {rgb}{0.03,0.03,0.03}
\definecolor {gray4}               {rgb}{0.04,0.04,0.04}
\definecolor {gray5}               {rgb}{0.05,0.05,0.05}
\definecolor {gray6}               {rgb}{0.06,0.06,0.06}
\definecolor {gray7}               {rgb}{0.07,0.07,0.07}
\definecolor {gray8}               {rgb}{0.08,0.08,0.08}
\definecolor {gray9}               {rgb}{0.09,0.09,0.09}
\definecolor {gray10}              {rgb}{0.10,0.10,0.10}
\definecolor {gray11}              {rgb}{0.11,0.11,0.11}
\definecolor {gray12}              {rgb}{0.12,0.12,0.12}
\definecolor {gray13}              {rgb}{0.13,0.13,0.13}
\definecolor {gray14}              {rgb}{0.14,0.14,0.14}
\definecolor {gray15}              {rgb}{0.15,0.15,0.15}
\definecolor {gray16}              {rgb}{0.16,0.16,0.16}
\definecolor {gray17}              {rgb}{0.17,0.17,0.17}
\definecolor {gray18}              {rgb}{0.18,0.18,0.18}
\definecolor {gray19}              {rgb}{0.19,0.19,0.19}
\definecolor {gray20}              {rgb}{0.20,0.20,0.20}
\definecolor {gray21}              {rgb}{0.21,0.21,0.21}
\definecolor {gray22}              {rgb}{0.22,0.22,0.22}
\definecolor {gray23}              {rgb}{0.23,0.23,0.23}
\definecolor {gray24}              {rgb}{0.24,0.24,0.24}
\definecolor {gray25}              {rgb}{0.25,0.25,0.25}
\definecolor {gray26}              {rgb}{0.26,0.26,0.26}
\definecolor {gray27}              {rgb}{0.27,0.27,0.27}
\definecolor {gray28}              {rgb}{0.28,0.28,0.28}
\definecolor {gray29}              {rgb}{0.29,0.29,0.29}
\definecolor {gray30}              {rgb}{0.30,0.30,0.30}
\definecolor {gray31}              {rgb}{0.31,0.31,0.31}
\definecolor {gray32}              {rgb}{0.32,0.32,0.32}
\definecolor {gray33}              {rgb}{0.33,0.33,0.33}
\definecolor {gray34}              {rgb}{0.34,0.34,0.34}
\definecolor {gray35}              {rgb}{0.35,0.35,0.35}
\definecolor {gray36}              {rgb}{0.36,0.36,0.36}
\definecolor {gray37}              {rgb}{0.37,0.37,0.37}
\definecolor {gray38}              {rgb}{0.38,0.38,0.38}
\definecolor {gray39}              {rgb}{0.39,0.39,0.39}
\definecolor {gray40}              {rgb}{0.40,0.40,0.40}
\definecolor {gray42}              {rgb}{0.42,0.42,0.42}
\definecolor {gray43}              {rgb}{0.43,0.43,0.43}
\definecolor {gray44}              {rgb}{0.44,0.44,0.44}
\definecolor {gray45}              {rgb}{0.45,0.45,0.45}
\definecolor {gray46}              {rgb}{0.46,0.46,0.46}
\definecolor {gray47}              {rgb}{0.47,0.47,0.47}
\definecolor {gray48}              {rgb}{0.48,0.48,0.48}
\definecolor {gray49}              {rgb}{0.49,0.49,0.49}
\definecolor {gray50}              {rgb}{0.50,0.50,0.50}
\definecolor {gray51}              {rgb}{0.51,0.51,0.51}
\definecolor {gray52}              {rgb}{0.52,0.52,0.52}
\definecolor {gray53}              {rgb}{0.53,0.53,0.53}
\definecolor {gray54}              {rgb}{0.54,0.54,0.54}
\definecolor {gray55}              {rgb}{0.55,0.55,0.55}
\definecolor {gray56}              {rgb}{0.56,0.56,0.56}
\definecolor {gray57}              {rgb}{0.57,0.57,0.57}
\definecolor {gray58}              {rgb}{0.58,0.58,0.58}
\definecolor {gray59}              {rgb}{0.59,0.59,0.59}
\definecolor {gray60}              {rgb}{0.60,0.60,0.60}
\definecolor {gray61}              {rgb}{0.61,0.61,0.61}
\definecolor {gray62}              {rgb}{0.62,0.62,0.62}
\definecolor {gray63}              {rgb}{0.63,0.63,0.63}
\definecolor {gray64}              {rgb}{0.64,0.64,0.64}
\definecolor {gray65}              {rgb}{0.65,0.65,0.65}
\definecolor {gray66}              {rgb}{0.66,0.66,0.66}
\definecolor {gray67}              {rgb}{0.67,0.67,0.67}
\definecolor {gray68}              {rgb}{0.68,0.68,0.68}
\definecolor {gray69}              {rgb}{0.69,0.69,0.69}
\definecolor {gray70}              {rgb}{0.70,0.70,0.70}
\definecolor {gray71}              {rgb}{0.71,0.71,0.71}
\definecolor {gray72}              {rgb}{0.72,0.72,0.72}
\definecolor {gray73}              {rgb}{0.73,0.73,0.73}
\definecolor {gray74}              {rgb}{0.74,0.74,0.74}
\definecolor {gray75}              {rgb}{0.75,0.75,0.75}
\definecolor {gray76}              {rgb}{0.76,0.76,0.76}
\definecolor {gray77}              {rgb}{0.77,0.77,0.77}
\definecolor {gray78}              {rgb}{0.78,0.78,0.78}
\definecolor {gray79}              {rgb}{0.79,0.79,0.79}
\definecolor {gray80}              {rgb}{0.80,0.80,0.80}
\definecolor {gray81}              {rgb}{0.81,0.81,0.81}
\definecolor {gray82}              {rgb}{0.82,0.82,0.82}
\definecolor {gray83}              {rgb}{0.83,0.83,0.83}
\definecolor {gray84}              {rgb}{0.84,0.84,0.84}
\definecolor {gray85}              {rgb}{0.85,0.85,0.85}
\definecolor {gray86}              {rgb}{0.86,0.86,0.86}
\definecolor {gray87}              {rgb}{0.87,0.87,0.87}
\definecolor {gray88}              {rgb}{0.88,0.88,0.88}
\definecolor {gray89}              {rgb}{0.89,0.89,0.89}
\definecolor {gray90}              {rgb}{0.90,0.90,0.90}
\definecolor {gray91}              {rgb}{0.91,0.91,0.91}
\definecolor {gray92}              {rgb}{0.92,0.92,0.92}
\definecolor {gray93}              {rgb}{0.93,0.93,0.93}
\definecolor {gray94}              {rgb}{0.94,0.94,0.94}
\definecolor {gray95}              {rgb}{0.95,0.95,0.95}
\definecolor {gray97}              {rgb}{0.97,0.97,0.97}
\definecolor {gray98}              {rgb}{0.98,0.98,0.98}
\definecolor {gray99}              {rgb}{0.99,0.99,0.99}
\definecolor {darkgrey}            {rgb}{0.66,0.66,0.66}

%% file: rs_macros_optsmt.tex
\newcommand{\omlarat}{\ensuremath{\text{OMT}(\larat)}\xspace}

\newcommand\mysout{\bgroup \markoverwith{{-}}\ULon}
\newcommand\nosout{\bgroup \markoverwith{{ }}\ULon}
\definecolor{mygray}{rgb}{0.90,0.90,0.90}
\definecolor{mywhite}{rgb}{1.00,1.00,1.00}

\newcommand{\optimathsat}{\textsc{OptiMathSAT}\xspace}

%% file: rs_macros_smt.tex
\newcommand{\muone}{\ensuremath{\mu_1}\xspace}
\newcommand{\mutwo}{\ensuremath{\mu_2}\xspace}

\newcommand{\smttt}[1]{\ensuremath{\text{SMT}(#1)}\xspace}

\newcommand{\larat}{\ensuremath{\mathcal{LA}(\mathbb{Q})}\xspace}

\renewcommand{\larat}{\ensuremath{\mathcal{LRA}}\xspace}

\newcommand{\smtlarat}{\smttt{\larat}}



%% file: rs_macros_specific.tex
\renewcommand{\larat}{\ensuremath{\mathcal{LRA}}\xspace}
\renewcommand{\smtlarat}{\ensuremath{\text{SMT}_{\larat}}\xspace}
\renewcommand{\smtlarat}{\smttt{\larat}}

\newcommand{\omttt}[1]{\ensuremath{\text{OMT}(#1)}\xspace}
\newcommand{\omtlarat}{\omttt{\larat}}

\renewcommand{\new}[1]{{\em #1}}

\newcommand{\preferred}[2]{\ensuremath{#1\succeq #2}}

\newcommand{\prefers}[2]{\ensuremath{ #1 \overset{\text{\tiny prefer}}{\longrightarrow} #2}}

\newcommand{\numunsatprefs}{\REQ{numUnsatPrefs}}
\newcommand{\numdeniedrequirements}{\REQ{numUnsatRequirements}}
\newcommand{\numsatisfiedtasks}{\REQ{numSatTasks}}

\newcommand{\calmone}{\ensuremath{\calm_1}\xspace}
\newcommand{\calmtwo}{\ensuremath{\calm_2}\xspace}
\newcommand{\mutwolex}{\ensuremath{\mu_2^{lex}}\xspace}
\newcommand{\mutwofam}{\ensuremath{\mu_2^{fam}}\xspace}
\newcommand{\mutwoeff}{\ensuremath{\mu_2^{eff}}\xspace}
\newcommand{\mutwopro}{\ensuremath{\mu_2^{pro}}\xspace}
\renewcommand{\mutwopro}{\ensuremath{\mu_{1\downarrow\calmtwo}}\xspace}
\renewcommand{\mutwopro}{\ensuremath{\mu_{1}}\xspace}

\newcommand{\familiaritycostof}[2]{\ensuremath{{\sf
FamiliarityCost(#1|#2)}\xspace}}
\newcommand{\effortcost}[2]{\ensuremath{{\sf ChangeEffort}(#1|#2)}\xspace}

\newcommand{\weightedfamiliaritycostof}[2]{\ensuremath{{\sf
WeightFamiliarityCost(#1|#2)}\xspace}}
\newcommand{\weightedeffortcostof}[2]{\ensuremath{{\sf WeightChangeEffort}(#1|#2)}\xspace}

%% file: mc_macros.tex
\usepackage{adjustbox}

\newcolumntype{R}[2]{%
    >{\adjustbox{angle=#1,lap=\width-(#2)}\bgroup}%
    l%
    <{\egroup}%
}

\tikzset{
  treenode/.style = {inner sep=0pt, text centered,
    font=\sffamily},
  arn_n/.style = {treenode, circle, black, font=\sffamily\bfseries, draw=black,
    text width=1.5em},
  arn_r/.style = {treenode, circle, black, font=\sffamily\bfseries, draw=black,
    text width=0.01em},
}

\newcommand{\REQ}[1]{\textsc{#1}}
\renewcommand{\REQ}[1]{\ensuremath{\mathsf{#1}}}

\newcommand{\refines}[2]{\ensuremath{ #1 \xrightarrow{R} #2}}
\newcommand{\Refines}[3]{\ensuremath{ #1 \xrightarrow{R_{#3}} #2}}
\newcommand{\contributes}[2]{\ensuremath{ #1 \xrightarrow{++} #2}}

\newcommand{\conflict}[2]{\ensuremath{ #1 \overset{--}{\longleftrightarrow} #2}}
\newcommand{\bind}[2]{\ensuremath{ #1{\longleftrightarrow} #2}}

\newcommand{\Goal}[1]{#1}

\newcommand{\Contribution}

\mathchardef\mhyphen="2D

\newcommand{\posConstraint}[2]{#2^{+}_{#1}}
\newcommand{\negConstraint}[2]{#2^{-}_{#1}}

\newcommand{\enum}[2]{#1_1, \ldots,#1_{#2}}

\newcommand{\Implies}{\rightarrow}
\newcommand{\equivalent}{\leftrightarrow}

\newenvironment{mcchange}{\color{darkgreen}}{\normalcolor}


%% file: abstract.tex
We are interested in supporting software evolution caused by changing 
requirements and/or changes in the operational environment of a 
software system. For example, users of a system may want new 
functionality or performance enhancements to cope with growing user 
population (changing requirements). Alternatively, vendors of a system 
may want to minimize costs in implementing requirements changes 
(evolution requirements). We propose to use Constrained Goal Models 
(CGMs) to represent the requirements of a system, and capture 
requirements changes in terms of incremental operations on a goal 
model. Evolution requirements are then represented as optimization 
goals that minimize implementation costs or customer value. We can then 
exploit reasoning techniques to derive optimal new specifications for an 
evolving software system. CGMs offer an expressive language for 
modelling goals that comes with scalable solvers that can solve hybrid 
constraint and optimization problems using a combination of Satisfiability 
Modulo Theories (SMT) and Optimization Modulo Theories (OMT) 
techniques. We evaluate our proposal by modeling and reasoning with a 
goal model for the meeting scheduling examplar.

%% file: intro.tex
We have come to live in a world where the only constant is
change. Changes need to be accommodated by any system that lives and
operates in that world, biological and/or engineered. For software
systems, this is a well-known problem referred to as software
evolution. There has been much work and interest on this problem since
Lehman's seminal proposal for laws of software evolution
\cite{Lehman80}. However, the problem of effectively supporting
software evolution through suitable concepts, tools and techniques is
still largely open. And software evolution still accounts for more
than 50\% of total costs in a software system's lifecycle. 

We are interested in supporting software evolution caused by changing
requirements and/or environmental conditions. Specifically, we are
interested in models that capture such changes, also in reasoning
techniques that derive optimal new specifications for a system whose
requirements and/or environment have changed. Moreover, we are
interested in discovering new classes of evolution requirements, in
the spirit of  
\cite{souza:phdthesis12} who proposed such a class for adaptive
software systems. We propose to model requirements changes through
changes to a goal model, and evolution requirements as optimization
goals, such as "Minimize costs while implementing new
functionality''. Our research baseline consists of an expressive
framework for modelling and reasoning with goals called Constrained
Goal Models (hereafter CGMs) \cite{nguyensgm16}. 
The CGM framework is founded on and
draws much of its power from Satisfiability Modulo Theories (SMT) and
Optimization Modulo Theories (OMT) solving techniques 
\cite{barrettsst09,sebastiani15_optimathsat}. 

The contributions of this paper include a proposal for modelling
changing requirements in terms of changes to a CGM model, but also the
identification of a new class of evolution requirements, expressed as
optimization goals in CGM. In addition, we show how to support
reasoning with changed goal models and evolution requirements in order
to derive optimal solutions. \ignoreinlong{\footnote{{\bf Note.}  
This paper was reduced to the
current size from its original 14-page length.
Accordingly, we have made available 
an extended  version of \cite{nguyensgm16_er16extended} 
including
{\em (i)} all figures of 
the examples which are described only verbally here, 
{\em (ii)} the {\em formalization} of the problem of
 automatically handling CGM evolutions and evolution requirements for
 CGMs,
{\em (iii)} an overview of our tool
implementing the presented approach,
{\em (iv)} an overview of related work, with a comparison 
wrt. previous approaches,
{\em (v)} some conclusions and description of future work.
}}

\ignoreinshort{
The rest of the paper is structured as follows:
\sref{sec:background} introduces the notion of CGM through a working example;
\sref{sec:evolving} introduces the notion of evolution requirements
and requirements evolution through our working example;
\ignoreinfinal{%
\sref{sec:evolutionreq_reasoning} formalizes the problem of
 automatically handling CGM evolutions and evolution requirements for CGMs;
\sref{sec:implementation} provides a brief overview of our tool
implementing the presented approach;
\sref{sec:related} overviews the related work,
and 
}
in \sref{sec:concl} we draw some conclusions and describe future work.
}

%% file: background.tex

%% file: background_ex.tex
\noindent {\bf \smtlarat and \omtlarat.} 
\new{Satisfiability Modulo the Theory of Linear Rational Arithmetic 
 (\smtlarat)} \cite{barrettsst09} is the problem of deciding the
satisfiability of arbitrary formulas on atomic propositions and
constraints in linear arithmetic over the rationals.
\new{Optimization Modulo the Theory of Linear Rational Arithmetic
(\omtlarat)} \cite{sebastiani15_optimathsat} extends \smtlarat by
searching solutions which optimize some \larat objective(s). 
Efficient \omlarat solvers like \optimathsat \cite{st_cav15} allow for
handling formulas with thousands of Boolean and rational variables
\cite{sebastiani15_optimathsat,nguyensgm16}.

\begin{figure*}
\centering 
\includegraphics[height=0.95\textheight, width=\textwidth]{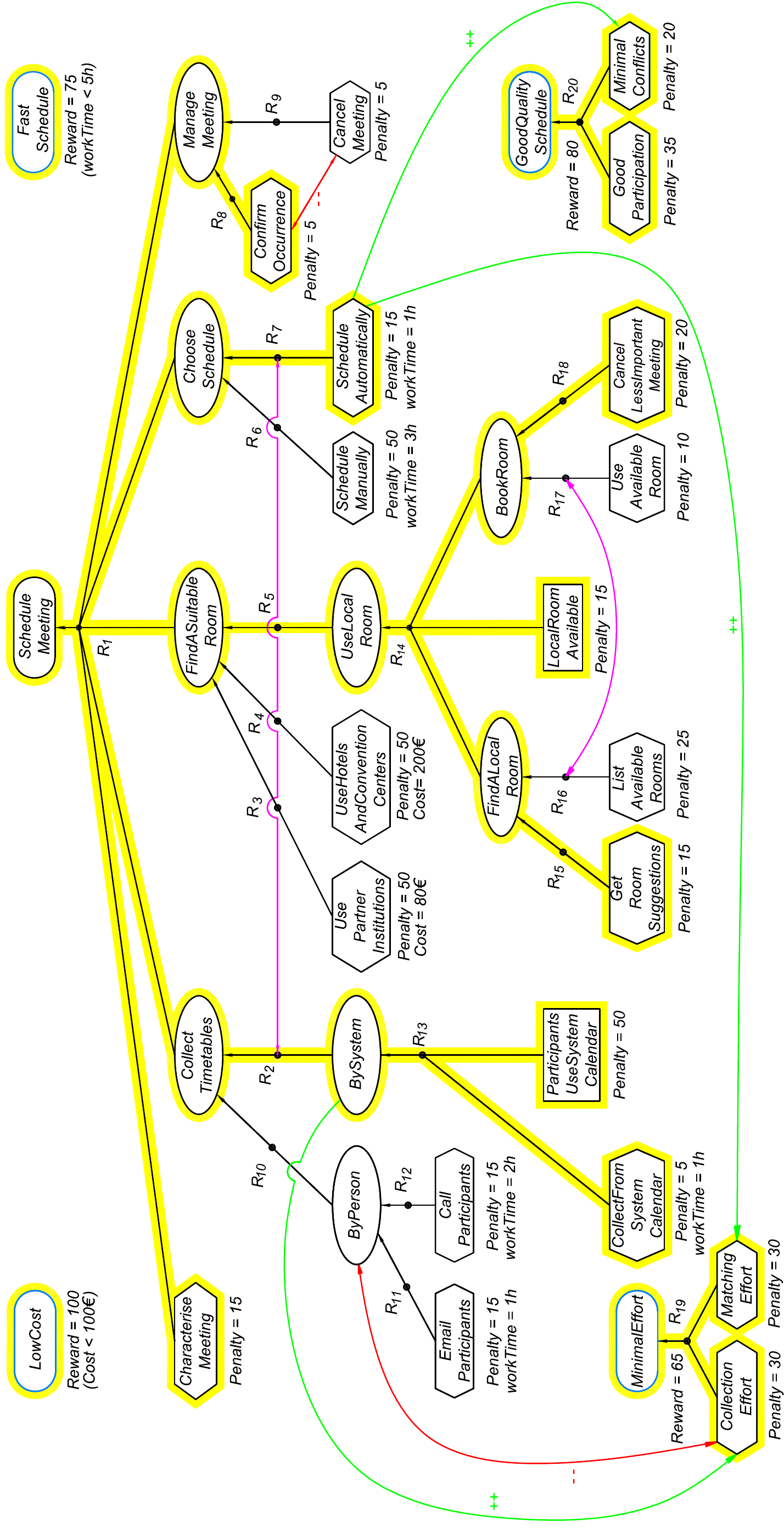}
\caption{{A CGM \calmone,  with a realization \muone minimizing
 lexicographically:
    the difference \REQ{Penalty}-\REQ{Reward}, \REQ{workTime}, 
and  \REQ{cost}.
Notationally, round-corner rectangles (e.g., \REQ{Schedule\-Meeting})
are root goals, representing stakeholder \new{requirements}; 
ovals (e.g. \REQ{Collect\-Timetables})
are \new{intermediate goals};
\new{hexagons} (e.g. \REQ{Characterise\-Meeting})
are \new{tasks}, i.e.  non-root leaf goals;
rectangles (e.g., \REQ{Participants\-Use\-System\-Calendar}) are
\new{domain assumptions}. 
\label{fig:CGMLex}
}
}
\end{figure*}

\noindent {\bf A Working Example.}
\label{sec:background_ex}
We recall from \cite{nguyensgm16} 
the main ideas of Constrained Goal Models (CGM's)
 and the main functionalities of
our CGM-Tool through a meeting scheduling example (Figure~\ref{fig:CGMLex}).
\ignore{
where we model the requirements for a meeting scheduling system, including
the functional requirement \REQ{Schedule\-Meeting}, 
as well as non-functional/quality requirements \REQ{Low\-Cost},  ~\REQ{Fast\-Schedule}, 
\REQ{Minimal\-Effort} and \REQ{Good\-Quality\-Schedule}. 
}
%
%
\ignoreinfinal{
Notationally, round-corner rectangles (e.g., \REQ{Schedule\-Meeting})
are root goals, representing stakeholder \new{requirements}; 
ovals (e.g. \REQ{Collect\-Timetables})
are \new{intermediate goals};
\new{hexagons} (e.g. \REQ{Characterise\-Meeting})
are \new{tasks}, i.e.  non-root leaf goals;
rectangles (e.g., \REQ{Participants\-Use\-System\-Calendar}) are
\new{domain assumptions}. 
}
We call \new{elements} both goals and domain assumptions.
Labeled bullets at the merging point of the edges 
connecting a group of source elements to a
target element   are \new{refinements}
(e.g., 
~\Refines{(\REQ{Good\-Participation},\REQ{Minimal\-Conflict})}{ \REQ{Good Quality
    Schedule}}{20}), while the $R_i$s denote
their labels.~%
{The label of a refinement can be omitted when
  there is no need to refer to it explicitly.}
 
%
Intuitively, requirements represent desired states of affairs we want the system-to-be
to achieve (either mandatorily or possibly); they are progressively
refined into intermediate goals, until the process produces actionable
goals (tasks) that need no further decomposition and can be executed;
domain assumptions are propositions about the domain that need to hold for a goal refinement to work.  {Refinements} are used to represent the alternatives of
how to achieve an element; a refinement of an element is a conjunction
of the sub-elements that are necessary to achieve it.

The main objective of the CGM in Figure~\ref{fig:CGMLex}
is to achieve the requirement \ignoreinlong{\break} \REQ{Schedule\-Meeting}, 
which is  \new{mandatory}. 
\REQ{Schedule\-Meeting} has only one candidate refinement $R_1$, 
consisting in five sub-goals: \REQ{Characterise\-Meeting}, \REQ{Collect\-Timetables}, 
\REQ{Find\-A\-Suitable\-Room}, \REQ{Choose\-Schedule}, and
~\REQ{Manage\-Meeting}. Since $R_1$ is the only refinement of the
requirement, all these sub-goals must be satisfied in order to
satisfy it. 
There may be more than one way to refine an element;
e.g., \REQ{Collect\-Timetables} is further refined either by
$R_{10}$ into the single goal \REQ{By\-Person} or by 
$R_{2}$ into the single goal \REQ{By\-System}.
\ignore{Similarly, \REQ{Find\-A\-Suitable\-Room} and 
\REQ{Choose\-Schedule} have three and two possible refinements respectively.}
The subgoals are further refined until they
reach the level of  domain assumptions and tasks.

Some requirements  
can be {``\new{nice-to-have}''}, 
like 
\REQ{Low\-Cost}, \REQ{Minimal\-Effort}, \ignoreinlong{\break} \REQ{Fast\-Schedule}, and
\REQ{Good\-Quality\-Schedule} (in blue in  Figure~\ref{fig:CGMLex}).
They are requirements that we would like to fulfill with our solution,
 provided they do not conflict with other requirements.
To this extent, in order to analyze interactively the 
possible different realizations, one can interactively mark [or unmark] 
requirements as satisfied, thus making them mandatory 
(if unmarked, they are nice-to-have ones). Similarly, 
one can interactively mark/unmark (effortful) tasks as denied,
or mark/unmark some domain assumption as satisfied or denied.
More generally, one can mark as satisfied or denied every goal or domain assumption.
We call these marks \new{user assertions}.


In a CGM, elements and refinements are enriched by user-defined
\new{constraints}, which can be expressed either graphically 
as  \new{relation edges} or textually as \new{Boolean or \ignoreinlong{\break}\smtlarat{} formulas}.
%
We have three kinds of relation edges.
\new{Contribution edges} \ignoreinlong{\break}``\contributes{E_i}{E_j}'' between elements  (in green in Figure~\ref{fig:CGMLex}),
like \ignoreinlong{\break}``\contributes{\REQ{Schedule\-Automatically}}{\REQ{Minimal\-Conflicts}}'', 
mean that if the source element $E_i$ is satisfied, then 
also the target element $E_j$ must be satisfied (but not vice versa). 
\new{Conflict edges} ``\conflict{E_i}{E_j}'' between elements (in red), like 
``\conflict{\REQ{Confirm\-Occurrence}}{\REQ{Cancel\-Meeting}}'',
mean that $E_i$ and $E_j$ cannot be both satisfied. 
\new{Refinement bindings} ``\bind{R_i}{R_j}'' between two refinements
(in purple), like ``\bind{R_2}{R_7}'', are used to state that, if the  target
elements $E_i$ and $E_j$ of the two refinements $R_i$ and $R_j$,
respectively, are both
satisfied, then $E_i$ is refined by $R_i$ if and only if $E_j$ is
refined by $R_j$. Intuitively, this means that the two refinements are
bound, as if they were two different instances of the same choice.

It is possible to enrich CGMs with logic formulas, representing
arbitrary logic 
constraints on elements and refinements.
%
\ignore{
E.g., a goal $\Goal{G}$  can be tagged with  
\new{prerequisite} formulas 
which 
must be
 satisfied  when  $\Goal{G}$ is
satisfied.
}
%
\ignoreinfinal{For example, to require that, as a prerequisite for \REQ{Fast\-Schedule},
\REQ{Schedule\-Manually} and \REQ{Call\-Participants} cannot be both satisfied, 
one can add the constraint 
"$\REQ{Fast\-Schedule}  \Implies  \neg (\REQ{Schedule\-Manually} \wedge \REQ{Call\-Participants})$''.}
In addition to Boolean constraints, 
it is also possible to use numerical variables to 
express different numerical attributes of elements 
(such as cost, worktime, space, fuel, etc.) and constraints over
them. For example, in Figure~\ref{fig:CGMLex} we 
associate to ~\REQ{Use\-Partner\-Institutions} and
~\REQ{Use\-Hotels\-And\-Convention\-Centers}
a cost value of 80\euro{} and 200\euro{} respectively, 
 and we associate ``$(\REQ{cost} < 100\euro{}) $'' as a
prerequisite constraint for the nice-to-have requirement
\REQ{Low\-Cost}. 
Implicitly, this means that no realization
involving \REQ{Use\-Hotels\-And\-Convention\-Centers} can realize this 
requirement.

%
We suppose now that \REQ{Schedule\-Meeting} is asserted as satisfied
(i.e. it is mandatory) and that no other element is asserted.
Then the CGM in Figure~\ref{fig:CGMLex} has more than 20 possible
\new{realizations}. The sub-graph which is 
highlighted in yellow describes one of them. 
Intuitively, a realization of a CGM under given user assertions (if
any) represents one of the alternative ways of refining the mandatory
requirements (plus possibly some of the nice-to-have ones) in compliance
with the user assertions and user-defined constraints.  It is a
sub-graph of the CGM including a set of satisfied elements and
refinements: it includes all mandatory requirements, and
[resp. does not include] all elements satisfied [resp. denied] in the
user assertions; for each non-leaf element included, at least one of
its refinement is included; for each refinement included, all its
target elements are included; finally, a realization complies with all
relation edges and with all constraints.

\ignore{
Apart from the mandatory requirement, 
the realization in Figure~\ref{fig:CGMLex} allows to achieve 
also the nice-to-have requirements \REQ{Low\-Cost},
\REQ{Good\-Quality\-Schedule}, but not \REQ{Fast\-Schedule} and \REQ{Minimal\-Effort};
it  requires accomplishing the tasks 
\REQ{Characterise\-Meeting}, 
\REQ{Call\-Participants},
\REQ{List\-Available\-Rooms},
\REQ{Use\-Available\-Room},
\REQ{Schedule\-Manually}, 
\REQ{Confirm\-Occurrence}, 
\REQ{Good\-Participation}, \REQ{Minimal\-Conflicts},
and requires  the domain assumption 
\REQ{Local\-Room\-Available}. 
}

\smallskip
In general, a CGM under given user assertions
has many possible realizations. To distinguish among them, 
stakeholders may want to express \new{preferences} on the requirements
 to achieve, on the tasks to accomplish, and on 
elements and refinements to choose. The CGM-Tool 
provides various methods to express preferences\ignoreinlong{, including}:
\begin{itemize}
\item 
attribute {\em rewards and penalties} to nice-to-have 
  requirements and tasks respectively, so that to maximize the
  former and minimize the latter; (E.g., satisfying 
  \REQ{Low\-Cost} gives a reward = 100, whilst satisfying 
\REQ{Characterise\-Meeting} gives a penalty = 15.)
\item 
introduce {\em numerical attributes}, 
  {\em constraints}  and  {\em objectives};
(E.g., the numerical attribute \REQ{Cost}  not only can be
used to set prerequisite constraints for requirements, 
like ``$(\REQ{Cost} < 100\euro{})$'' for \REQ{LowCost}, but also can be
set as objectives to minimize.)
\ignoreinfinal{
\item introduce a list of {\em binary preference relations}
  ``\preferred{}{}''between elements or  refinements.
(E.g., one can set the preferences
$\preferred{\REQ{By\-System}}{\REQ{By\-Person}}$, 
$\preferred{\REQ{Use\-Local\-Room}}{\REQ{Use\-Partner\-Institutions}}$ and
$\preferred{\REQ{Use\-Local\-Room}}{\REQ{Use\-Hotels\-And\-Convention\-Centers}}$.)
}
\end{itemize}
\noindent
\ignoreinlong{
The CGM-Tool provides many automated-reasoning functionalities on CGMs \cite{nguyensgm16}.
\begin{itemize} 
\item[{\em Search/enumerate minimum-penalty/maximum reward realizations.}]
  One
  can assert rewards to the desired requirements and set penalties
  of tasks, then the tool 
  finds automatically the optimal realization(s).
\item[{\em Search/enumerate optimal realizations wrt. pre-defined/user-defined
  objectives.}] One \ignoreinlong{\break} can define objective functions
  $obj_1,...,obj_k$ over goals, refinements and their numerical
  attributes; then the tool finds automatically realizations optimizing
  them. 
\end{itemize}
}
\ignoreinfinal{
The CGM-Tool provides many automated-reasoning functionalities on CGMs~\cite{nguyensgm16}.
\begin{itemize} 
\item[{\em Search/enumerate realizations.}] 
  One can automatically check the realizability of a CGM--or to
  enumerate one or more of its possible realizations-- under a group
  of user assertions and of user-defined 
  constraints.
(When a CGM is found un-realizable under a group of user
assertions and of user-defined 
  constraints, it highlights the
subparts of the CGM and the subset of assertions causing the
problem.)
\item[{\em Search/enumerate minimum-penalty/maximum reward realizations.}]
  One
  can assert rewards to the desired requirements and set penalties
  of tasks, then the tool 
  finds automatically the optimal realization(s).
\item[{\em Search/enumerate optimal realizations wrt. pre-defined/user-defined
  objectives.}] One can define objective functions
  $obj_1,...,obj_k$ over goals, refinements and their numerical
  attributes; then the tool finds automatically realizations optimizing
  them. 
\item[{\em Search/enumerate optimal realizations wrt. binary preferences.}] 
Once the list of binary preference is set, the tool  finds
automatically realizations maximizing the number of fulfilled preferences.
\end{itemize}
}
\noi
The above functionalities can be combined in various ways. 
For instance, the realization of
Figure~\ref{fig:CGMLex} is the one returned by CGM-tool 
when asked to  minimize lexicographically, in order,  the difference 
\REQ{Penalty}-\REQ{Reward},  \REQ{workTime}, 
and  \REQ{cost}.~\footnote{{A solution \new{optimizes
    lexicographically} an ordered list of objectives
    \tuple{obj_1,obj_2,...} if it makes $obj_1$ optimum
  and, if more than one such solution exists, it makes also $obj_2$
  optimum, ..., etc. }}
They have been implemented by encoding the CGM
and the objectives into an \smtlarat formula
 and a set of \larat objectives, which is fed to
the OMT tool \optimathsat \cite{st_cav15}.
We refer the reader to  \cite{nguyensgm16} for a much more detailed 
description of CGMs and their automated reasoning functionalities.


%% file: evolving.tex
\ignoreinshort{Here we show how a CGM can evolve, and how we can handle such evolution.}

%% file: reqevolution.tex
\JMSUGGESTION{{Section 3.1 would describe how we model changes to requirements,
ie some goals are added/removed.}\\}

\noi
\ignoreinlong{\bf{Requirements Evolution.}} Constrained goal models may evolve in time: goals, requirements and
assumptions can be added, removed, or simply modified; 
Boolean and SMT constraints may be added, removed, or modified as
well;
assumptions which were assumed true can be assumed false, or vice
versa.

Some modifications {\em strengthen} the CGMs, in the sense that they
reduce the set of candidate realizations. For instance, dropping one
of the refinements of an element (if at least one is left) reduces the
alternatives in realizations; adding source elements to a refinement
makes it harder to satisfy; adding Boolean or SMT constraints, or
making some such constraint strictly stronger, restricts the set of candidate
solutions; changing the value of an assumption from true to false may
drop some alternative solutions.
Vice versa, some modifications {\em weaken} the CGMs, augmenting the
set of candidate realizations: for instance, adding one of refinement
to an element, dropping source elements to a refinement, dropping
Boolean or SMT constraints, or making some such constraint strictly
weaker, changing the value of an assumption from false to true.
In general, however, since in a CGM the goal and/or decomposition graph is a DAG
and not a tree, and the and/or decomposition is augmented with
relational edges and constraints, modifications may produce
combinations of the above effects, possibly propagating unexpected
side effects which are sometimes hard to predict.



We consider the CGM \ignoreinshort{of a Schedule Meeting described} in
Figure~\ref{fig:CGMLex}  (namely, $\calm_1$) as our starting model, and 
we assume that for some reasons it has been modified into the CGM $\calm_2$
\ignoreinfinal{of Figure~\ref{fig:mu2pro}.}
\ignoreinlong{of Figure~2 in \cite{nguyensgm16_er16extended} (see \sref{sec:intro}).}
$\calm_2$ differs from $\calm_1$ 
for the following modifications:
\begin{aenumerate}
\item \label{item:R13}
two new tasks, \REQ{Set System Calendar} and
\REQ{Participants Fill System Calendar}, are added to the sub-goal
sources of the refinement $R_{13}$;
\item \label{item:R17} a new source task \REQ{RegisterMeetingRoom} is
  added to {$R_{17}$}, and 
the binding between {$R_{16}$} and {$R_{17}$} is removed;
the refinement $R_{18}$ of the goal
\REQ{BookRoom} and its source task
\REQ{CancelLessImportantMeeting} are removed;
\item 
the alternative
refinements $R_8$ and $R_9$ of \REQ{Manage Meeting} are also modified:
two new internal goals \REQ{By User} and \REQ{By Agent} are added and
become the single source of the two refinements $R_8$ and $R_9$
respectively, and the two tasks \REQ{Confirm Occurrence} and \REQ{Cancel
  Meeting} become respectively the sources of two new refinements
$R_{21}$ and $R_{22}$, 
which are the alternative refinements of the goal \REQ{By User};
the
new goal \REQ{By Agent} is refined by the new refinement
$R_{23}$ with source  task \REQ{Send Decision}.
\end{aenumerate}

\ignoreinfinal{
\begin{figure*}
\centering 
\includegraphics[angle=90,origin=c,height=0.75\textheight, width=.9\textwidth]{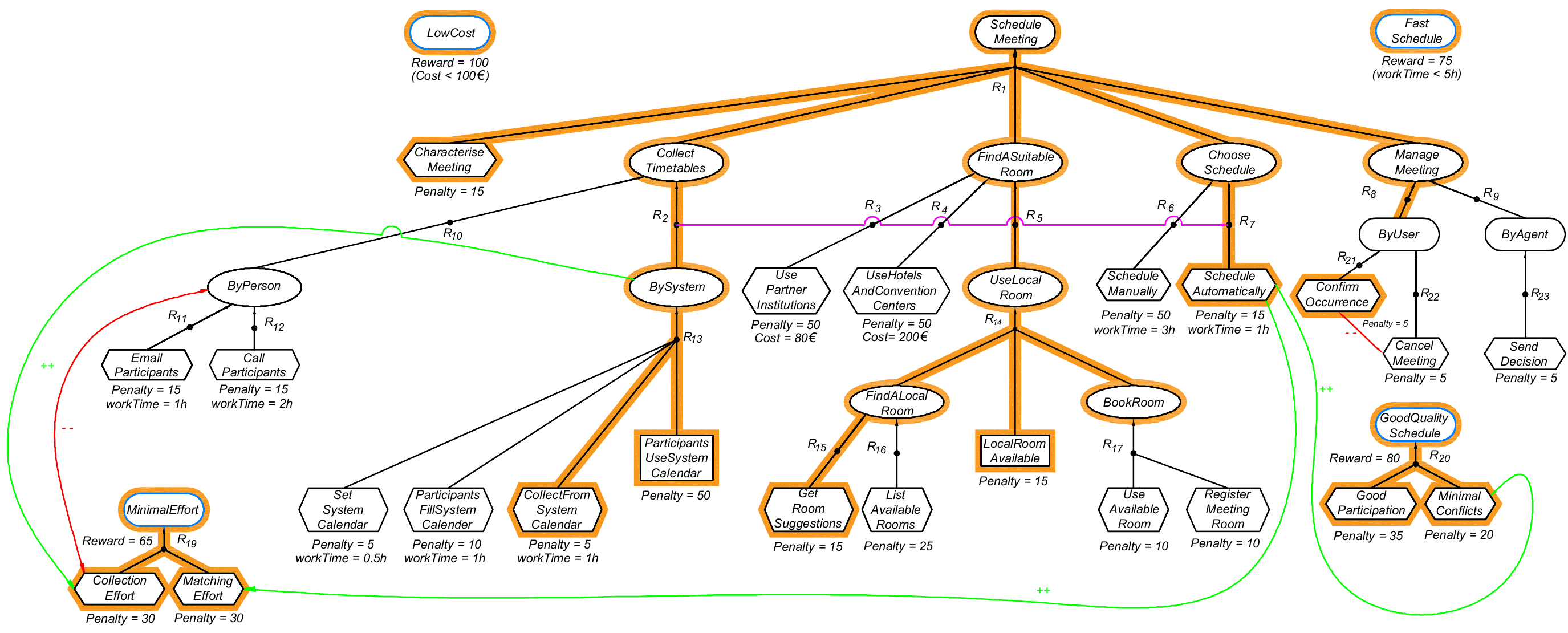}
\caption{{The novel CGM \calmtwo, with the previous realization \mutwopro
    highlighted  for comparison. (Notice that \muone is no more a
    valid realization for \calmtwo.)}
\label{fig:mu2pro}}
\end{figure*}
}

%% file: evolutionreq.tex
\JMSUGGESTION{Section 3.2 would describe requirements on
the evolution of a system. This is where we show how to formalize max
familiarity and min effort. These ARE requirements on the evolution of the
system, after all.}

\noindent\ignoreinlong{\bf{Evolution Requirements.}} We consider the generic scenario in which a previous version of a CGM
\calmone with an available realization \muone 
is modified into a new CGM \calmtwo.  
%
%
As a consequence\ignoreinfinal{\ of modifying a CGM \calmone into a new version
\calmtwo},  \mutwopro
typically is no more a valid realization of \calmtwo.\ignoreinfinal{~\footnote{
More precisely, rather than ``\muone'', here we should say 
 ``the restriction of \muone to the
elements and variables which are still in \calmtwo.'' 
We will keep this distinction implicit in the rest of the paper.}}
E.g., we notice that \mutwopro 
\ignoreinfinal{in Figure~\ref{fig:mu2pro}}
\ignoreinlong{in Figure~2 in \cite{nguyensgm16_er16extended}}
does not
represent a valid realization of \calmtwo: not all source tasks
of $R_{13}$ are satisfied, \REQ{BookRoom} has no satisfied refinement,
and the new goal \REQ{ByUser} and refinement $R_{21}$ are not
satisfied.
It is thus necessary to produce a new realization \mutwo for
\calmtwo. 

In general, when one has a sequence $\calmone,\calmtwo,...,\calm_i,...$
of CGMs and must produce a corresponding sequence
$\muone,\mutwo,...,\mu_i,...$ of realizations, it is necessary to decide some
criteria by which the realizations $\mu_i$ evolve in terms of 
the evolution of the CGMs $\calm_i$. We call these criteria, \new{evolution
requirements}.
We describe some possible criteria. 

\noindent {\bf Recomputing realizations.}
One possible evolution requirement is that of always having  the ``best'' 
realization $\mu_i$ for each $\calm_i$, according to some 
objective (or lexicographic combination of objectives).
Let \calmone, \calmtwo, and \muone  be as above. 
One possible choice  for the user is to compute a new optimal
realization \mutwo from scratch, using the same criteria used in
computing \muone from \calmone. 
In general, however, it may be the case that the new realization \mutwo is
very different from \muone, which may displease the stakeholders.

We consider now the realization $\muone$ of the CGM \calmone
highlighted in Figure~\ref{fig:CGMLex} and the modified model \calmtwo
of 
\ignoreinfinal{Figure~\ref{fig:mu2pro}.}  
\ignoreinlong{Figure~2 in \cite{nguyensgm16_er16extended}.}  
If we run CGM-Tool over \calmtwo 
with the same optimization criteria as for $\muone$ --i.e., 
minimize lexicographically, in order,
    the difference \REQ{Penalty}-\REQ{Reward}, \REQ{workTime}, 
and  \REQ{cost}-- we obtain a novel realization \mutwolex{} 
\ignoreinshort{depicted in Figure~\ref{fig:mu2lex}.}
\ignoreinlong{(Figure~3 in \cite{nguyensgm16_er16extended}.}
The new realization \mutwolex{} satisfies all the requirements (both "nice to have" and mandatory) except \REQ{MinimalEffort}. It includes the following tasks:
\REQ{CharateriseMeeting},
\REQ{EmailParticipants},
\REQ{GetRoomSuggestions},
\REQ{UseAvailableRoom},
\ignoreinlong{\break}\REQ{RegisterMeetingRoom},
\REQ{ScheduleManually},
\REQ{ConfirmOccurrence},
\REQ{GoodParticipation}, \ignoreinlong{\break} and
\REQ{MinimalConflicts}, 
and it requires one domain assumption: \REQ{LocalRoomAvailable}. This realization was found automatically
by our CGM-Tool in 0.059 seconds on an Apple MacBook Air laptop.

Unfortunately, \mutwolex turns out to be extremely different from 
\muone. This is due to the fact that the novel tasks \REQ{SetSystemCalendar} 
and \REQ{ParticipantsFillSystemCalendar} raise significantly the penalty
for $R_{13}$ and thus for $R_{2}$; hence, in terms of the
\REQ{Penalty}-\REQ{Reward} objective, it is now better to 
choose $R_{10}$ and $R_{6}$ instead of $R_{2}$ and $R_{7}$, even
though this forces \REQ{ByPerson} to be satisfied, which is
incompatible with \REQ{CollectionEffort}, so that \REQ{MinimalEffort} 
is no more achieved. 
Overall, for $\mutwo$ we have 
$\REQ{Penalty}-\REQ{Reward} = -65$, $\REQ{workTime} = 4h$
and $\REQ{cost} = 0\euro$.


In many contexts, in particular if \muone is well-established
or is already implemented, one may want to find a realization
\mutwo of the modified CGM \calmtwo which is as similar as possible to
the previous realization \calmone.
The suitable notion of "similarity'', however, may depend on
stakeholder's needs.
%
%
%
In what follows, we discuss two notions of "similarity'' from
\cite{ErnstBMJ12}, {\em familiarity} and  
{\em change effort}, adapting and extending them to CGMs.

\noindent {\bf Maximizing familiarity.}
%
%
In our approach, in its simplest form, the \new{familiarity} of \mutwo
wrt. \muone is given by the number of elements of interest 
 which are common to \calmone and \calmtwo and which either
are in both $\mu_1$ and $\mu_2$ or are out of both of them; this can be
augmented also by the number of new elements in \calmtwo of interest
(e.g., tasks) which are denied. 
In a more sophisticate form, the contribution of each element of
interest can be weighted by some numerical value 
 (e.g., \REQ{Penalty}, \REQ{cost}, \REQ{WorkTime},...).
\ignoreinfinal{This is formalized in
\sref{sec:evolutionreq_reasoning}, and a functionality for maximizing
familiarity is implemented in CGM-Tool.}

For example, if we ask CGM-Tool to find a realization which maximizes our 
notion of familiarity\ignoreinfinal{ (see \sref{sec:evolutionreq_reasoning})}, we
obtain the novel realization \mutwofam 
\ignoreinshort{depicted in Figure~\ref{fig:mu2fam}.}
\ignoreinlong{(Figure~4
 in \cite{nguyensgm16_er16extended}).}
\mutwofam satisfies all the requirements (both "nice to have" and mandatory ones), and includes the following tasks: 
\REQ{CharacteriseMeeitng},
\REQ{SetSystemCalendar},
\REQ{ParticipantsFillSystemCalendar},
\REQ{CollectFromSystemCalendar},
\REQ{GetRoomSuggestions},
\REQ{UseAvailableRoom},
\ignoreinlong{\break}\REQ{RegisterMeetingRoom},
\REQ{ScheduleAutomatically},
\REQ{ConfirmOccurrence},
\ignoreinlong{\break}\REQ{GoodParticipation}, 
\REQ{MinimalConflicts},
\REQ{CollectionEffort}, and
\REQ{MatchingEffort};
$\mutwofam$ also requires two domain assumptions: 
\REQ{ParticipantsUseSystemCalendar} and 
\ignoreinlong{\break}\REQ{LocalRoomAvailable}.

Notice that 
all the tasks
which are satisfied 
in \mutwopro are satisfied also in $\mutwofam$, and
only the intermediate goal \REQ{ByUser}, the refinement $R_{21}$ and the
four tasks 
\REQ{SetSystemCalendar},
\REQ{ParticipantsFillSystemCalendar},
\REQ{UseAvailableRoom}, and 
\REQ{RegisterMeetingRoom} are added to \mutwofam,
three of which are newly-added tasks. 
Thus, on common elements, \mutwofam and \mutwopro differ only on the task
\REQ{UseAvailableRoom}, which
must be mandatorily be satisfied to complete the realization. 
Overall, wrt. \mutwolex, we pay familiarity with some loss 
in the ``quality'' of the realization, since for $\mutwofam$ we have 
$\REQ{Penalty}-\REQ{Reward} = -50$, $\REQ{workTime} = 3.5h$
and $\REQ{cost} = 0\euro$.  
This realization was found automatically
by our CGM-Tool in 0.067 seconds on an Apple MacBook Air laptop.
%

\noindent {\bf Minimizing change effort.}
In our approach, in its simplest form, 
the \new{change effort} of \mutwo wrt. \muone is
given by the number of newly-satisfied tasks, i.e., the amount of the
new tasks which are satisfied in \mutwo plus that of common
tasks which were not satisfied in \muone but are satisfied in \mutwo. 
In a more sophisticate form, the contribution of each task of interest
can be weighted by some numerical
value (e.g., \REQ{Penalty}, \REQ{cost}, \REQ{WorkTime},...).
Intuitively, since satisfying a task requires effort, this value 
considers the extra effort required to implement \mutwo. 
(Notice that tasks which pass from satisfied to denied do not 
reduce the effort, because we assume they have been implemented anyway.)
\ignoreinfinal{This is formalized in \sref{sec:evolutionreq_reasoning}, and 
a functionality for minimizing change effort is
implemented in CGM-Tool.}

For example, if we ask CGM-Tool to find a realization which minimizes 
the number of newly-satisfied tasks, we
obtain the realization $\mutwoeff$ 
\ignoreinshort{depicted in Figure~\ref{fig:mu2eff}.}
\ignoreinlong{(Figure~5 in \cite{nguyensgm16_er16extended}).}
{The realization satisfies all the requirements (both "nice to have" and mandatory), and includes the following tasks: 
\REQ{CharacteriseMeeitng},
\REQ{SetSystemCalendar},
\ignoreinlong{\break}\REQ{ParticipantsFillSystemCalendar},
\REQ{CollectFromSystemCalendar},
\REQ{UsePartnerInstitutions},
\REQ{ScheduleAutomatically},
\REQ{ConfirmOccurrence},
\REQ{GoodParticipation}, 
\REQ{MinimalConflicts},
\ignoreinlong{\break}\REQ{CollectionEffort}, and
\REQ{MatchingEffort};
$\mutwoeff$ also requires one domain assumption
\ignoreinlong{\break}\REQ{ParticipantsUseSystemCalendar}.

Notice that, in order to minimize the number of new tasks needed to be achieved, 
in $\mutwoeff$, \REQ{FindASuitableRoom} is refined by $R_3$ instead of $R_5$. 
In fact, in order to achieve $R_5$, we would need to satisfy two extra tasks 
(\REQ{UseAvailableRoom} and \ignoreinlong{\break}\REQ{RegisterMeetingRoom}) wrt. $\mu_1$, whilst for satisfying $R_3$ we only need to satisfy one task
(\REQ{UsePartnerInstitutions}). Besides, two newly added tasks 
\REQ{SetSystemCalendar} and
\REQ{ParticipantsFillSystemCalendar}
are also included in $\mutwoeff$. Thus the total effort of evolving from $\mu_1$
to $\mutwoeff$ is to implement
three new tasks.
Overall, for $\mutwoeff$ we have 
$\REQ{Penalty}-\REQ{Reward} = -50$, $\REQ{workTime} = 3.5h$
and $\REQ{cost} = 80\euro$.  
This realization was found automatically
by our CGM-Tool in 0.085 seconds on an Apple MacBook Air laptop.



\noindent {\bf Combining familiarity or change effort
 with other objectives.} In our approach,  familiarity and change
effort are numerical objectives like others, and as such they 
 can be combined lexicographically with
other objectives, so that stakeholders can decide which objectives to
prioritize.

%% file: appendix.tex
\begin{figure*}
\centering 
\includegraphics[angle=90,origin=c,height=0.75\textheight, width=.9\textwidth]{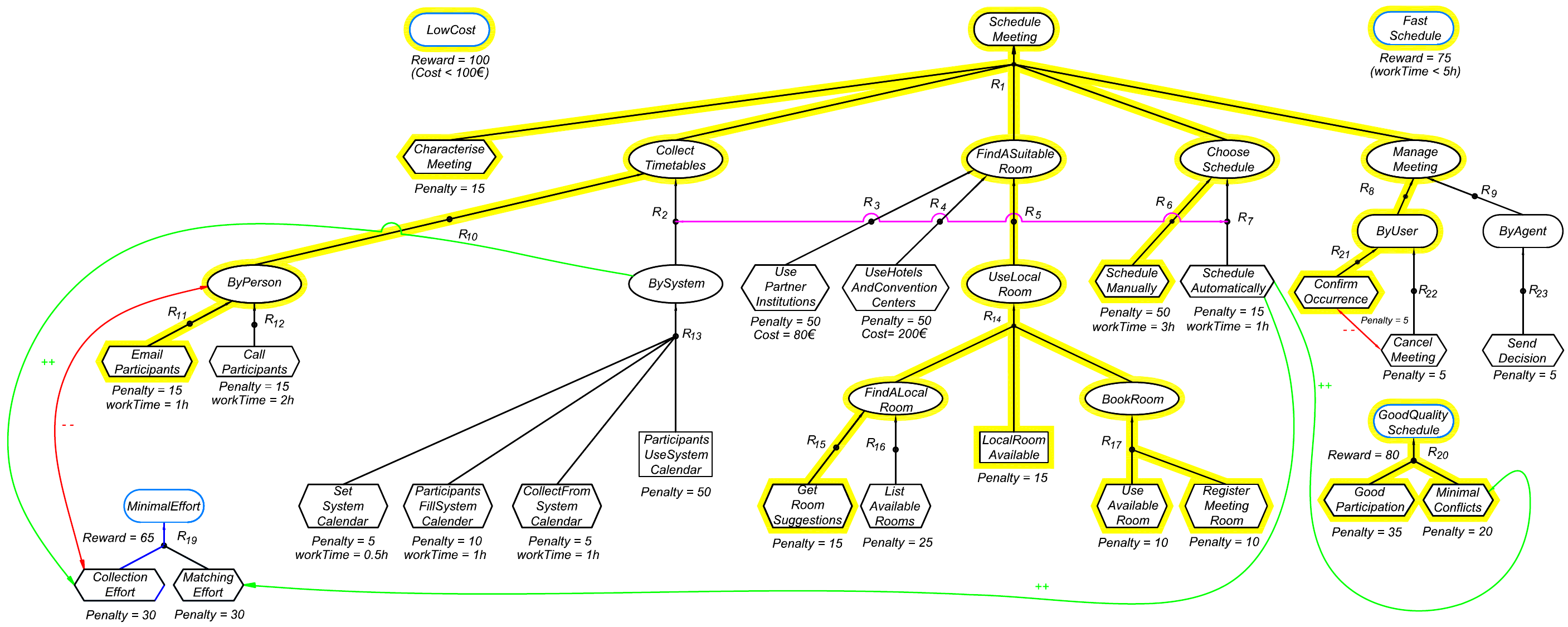}
\caption{{New CGM \calmtwo, with realization \mutwolex which minimizes
    lexicographically: the difference \REQ{Penalty}-\REQ{Reward},
    \REQ{workTime}, and  \REQ{cost}.}
\label{fig:mu2lex}}
\end{figure*}

\begin{figure*}
\centering 
\includegraphics[angle=90,origin=c,height=0.75\textheight, width=.9\textwidth]{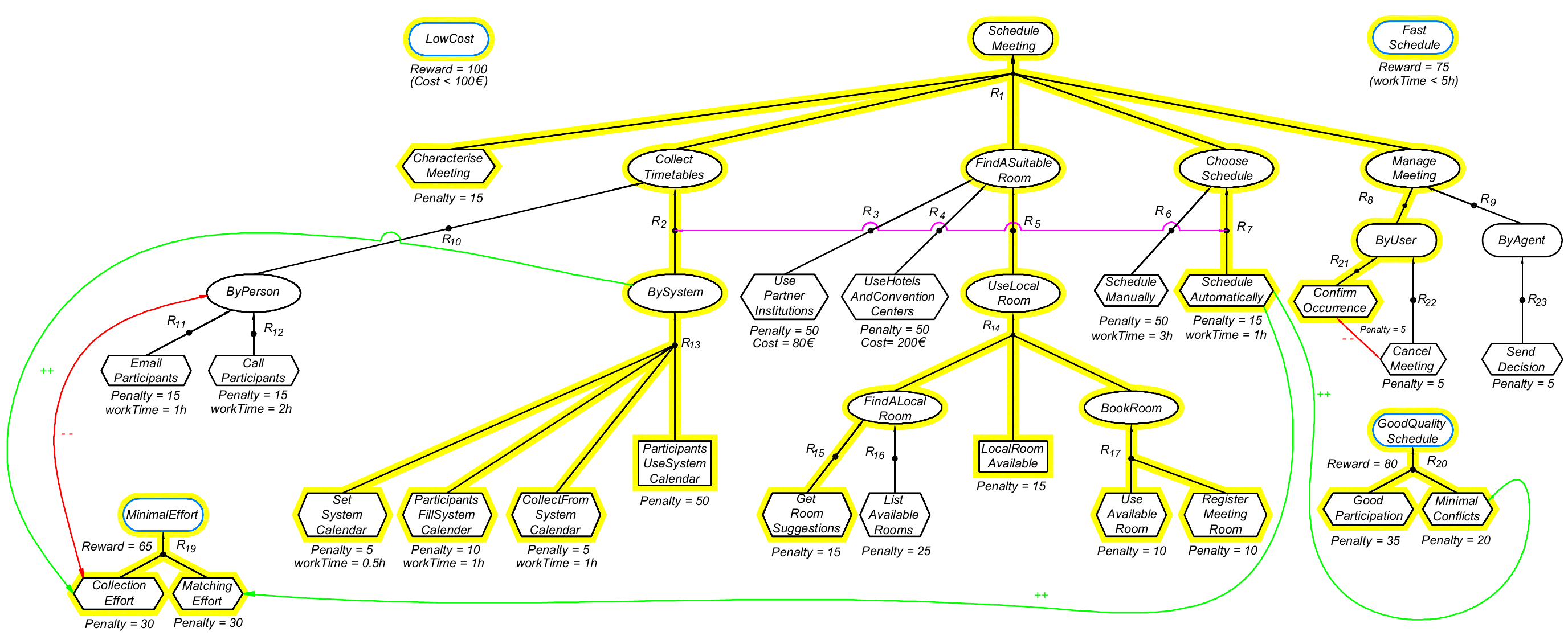}
\caption{{New CGM \calmtwo, with realization \mutwofam with maximizes
    the familiarity wrt. \muone.}
\label{fig:mu2fam}}
\end{figure*}

\begin{figure*}
\centering 
\includegraphics[angle=90,origin=c,height=0.75\textheight, width=.9\textwidth]{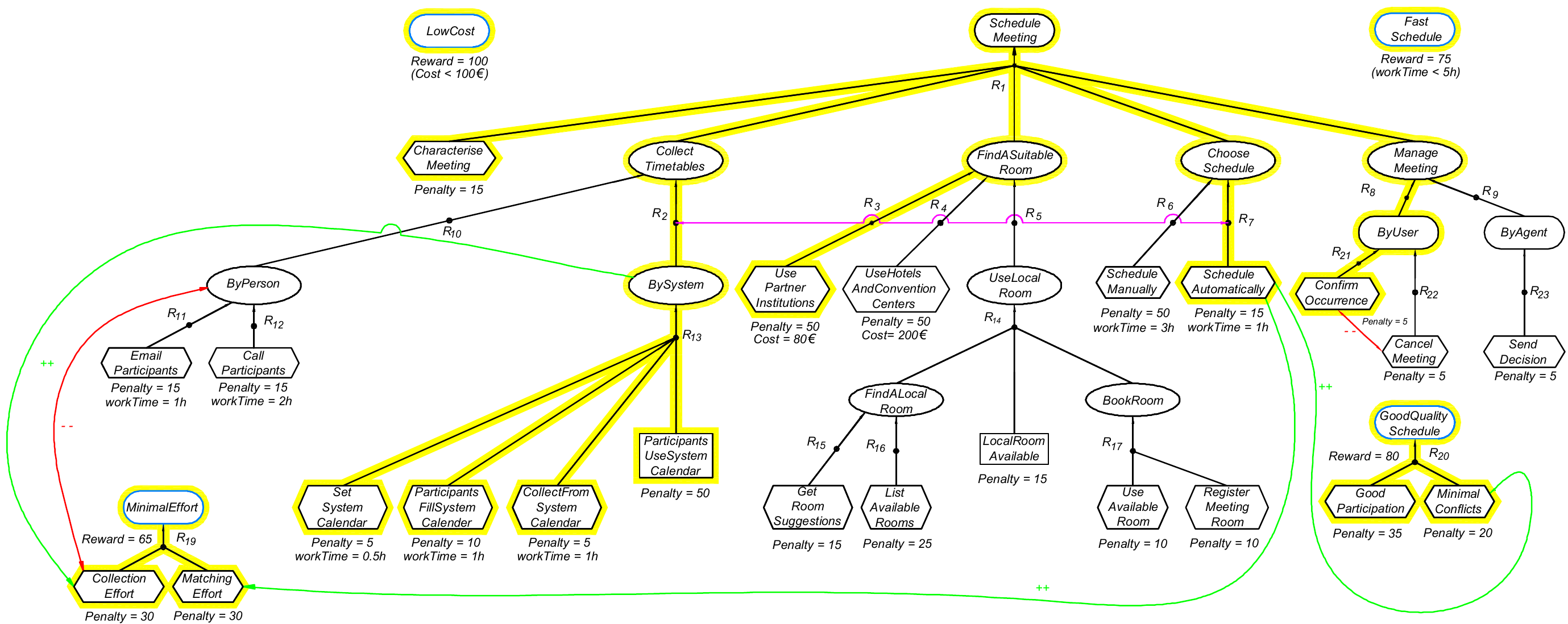}
\caption{{New CGM \calmtwo, with realization \mutwoeff with minimimizes
    the change effort wrt. \muone.}
\label{fig:mu2eff}}
\end{figure*}

\ignore{
\begin{figure*}
\centering 
\includegraphics[angle=90,origin=c,height=0.75\textheight, width=.9\textwidth]{}
\caption{{}
\label{fig:}}
\end{figure*}
}

\ignore{

\begin{figure*}
\centering 
\includegraphics[angle=90,origin=c,height=0.75\textheight, width=.9\textwidth]{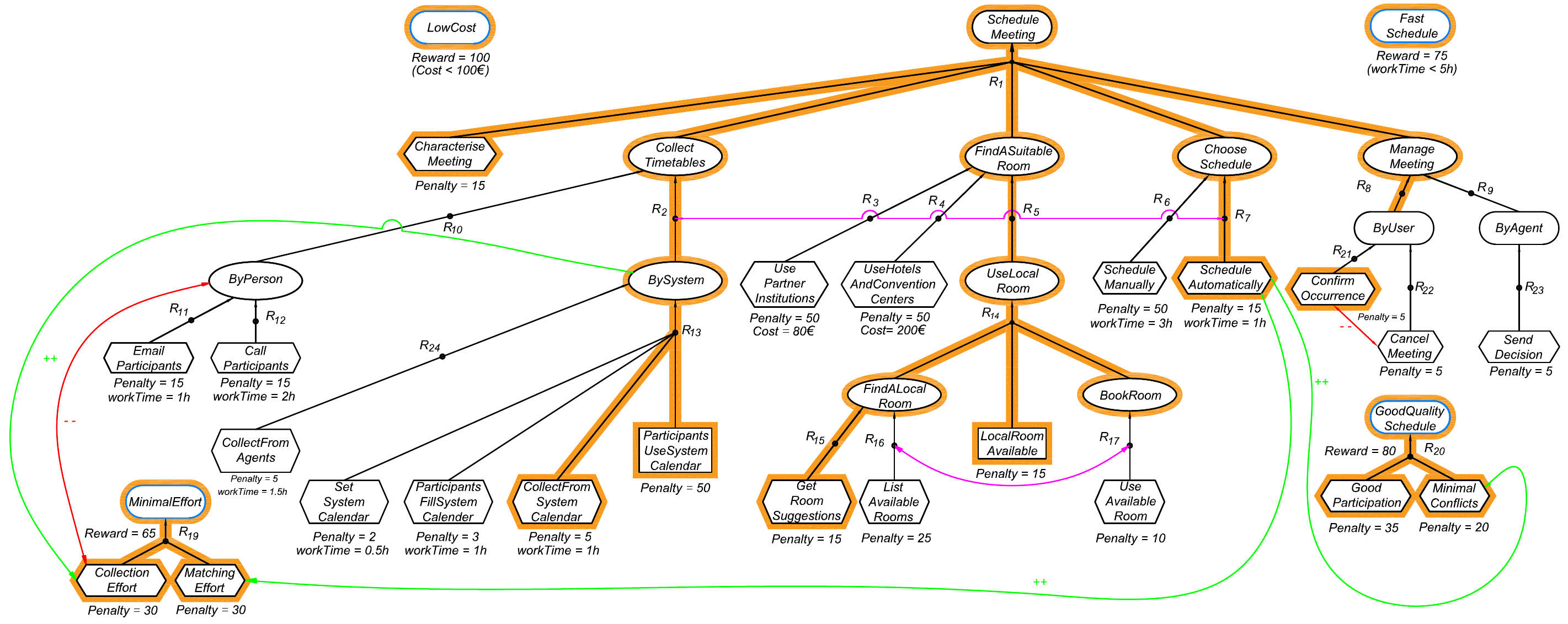}
\caption{{Restriction}
\label{fig:CGMEv}}
\end{figure*}

\subsection*{OLD ONES}
\begin{figure*}
\centering 
\includegraphics[angle=90,origin=c,height=0.75\textheight, width=.9\textwidth]{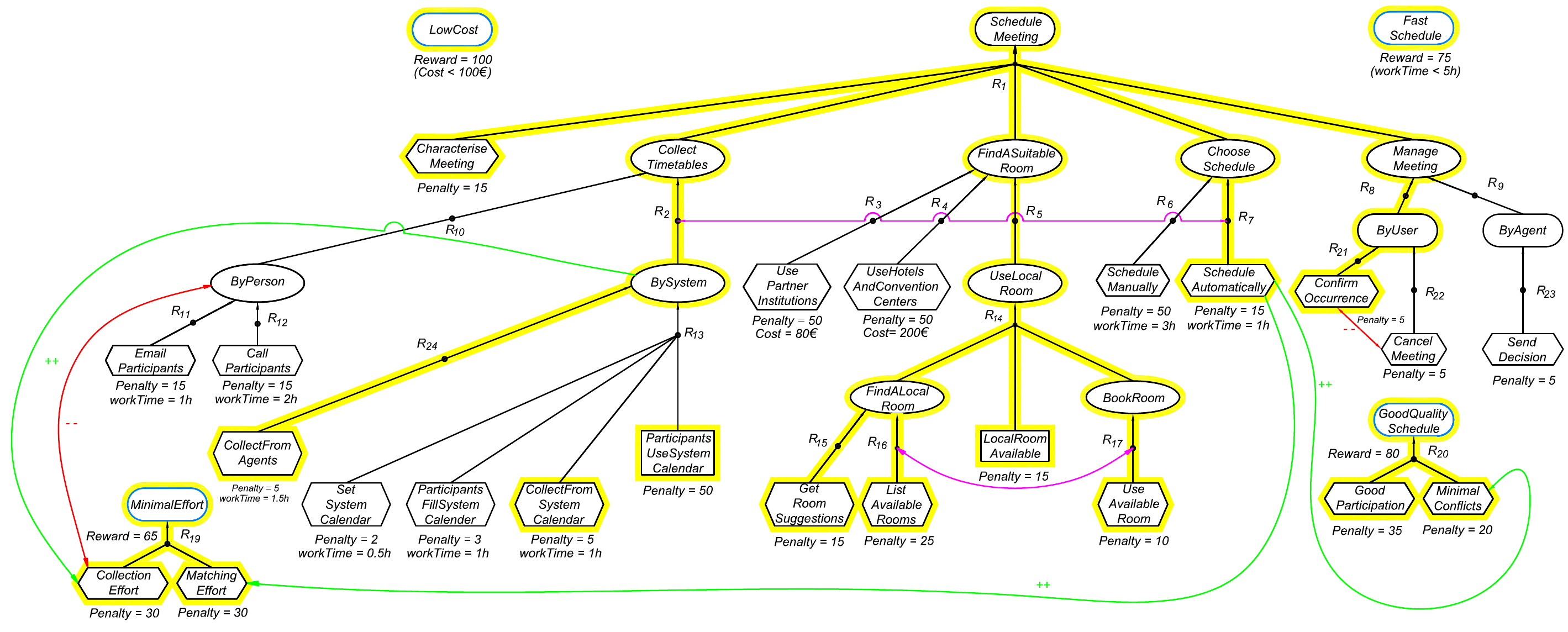}
\caption{{Familiarity.}
\label{fig:CGMFamiliarity}}
\end{figure*}
\begin{figure*}
\centering 
\includegraphics[angle=90,origin=c,height=0.75\textheight, width=.9\textwidth]{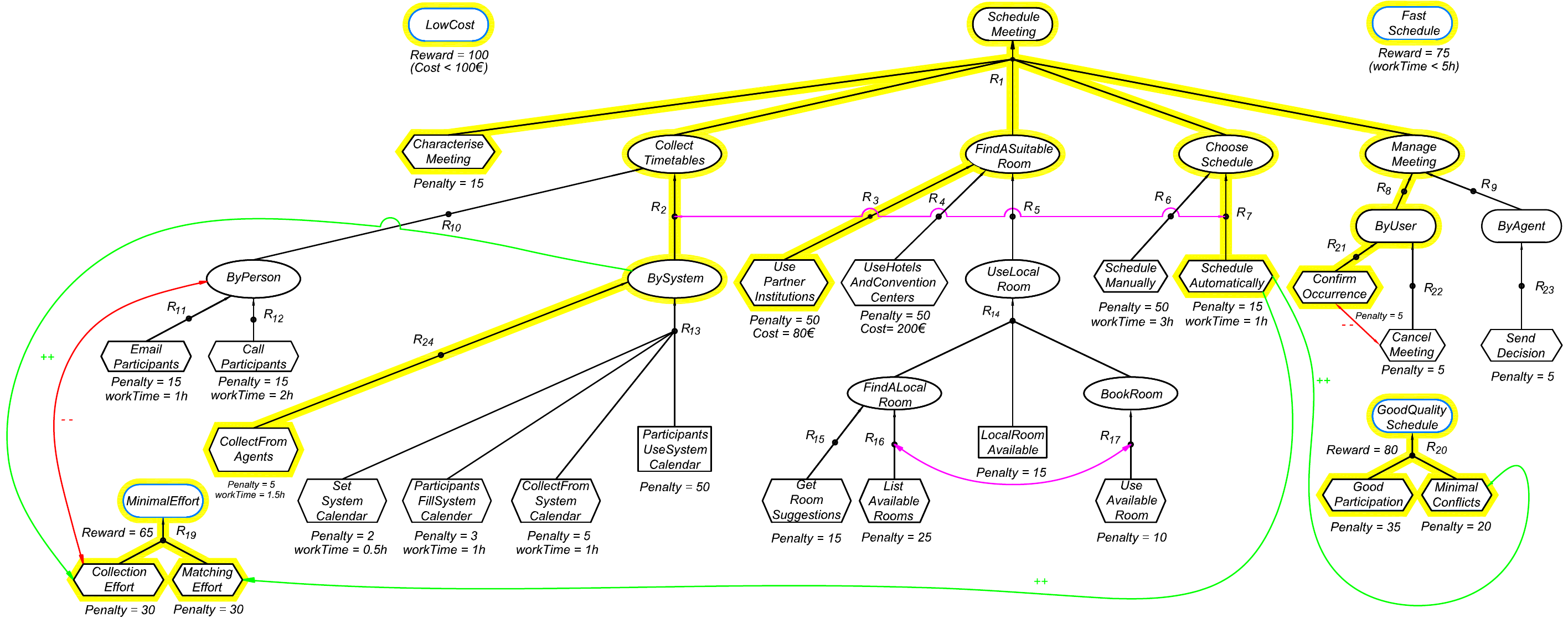}
\caption{Effort.
\label{fig:CGMEffort}}
\end{figure*}

}

%% file: background_formal.tex
\fakesubsubsection{CGMs and realizations}
We first recall some formal definitions from  \cite{nguyensgm16}.

\label{def:cgm}
A \new{Constrained Goal Model (CGM)} is a tuple $\calm\defas\tuple{\calb,\caln,\cald,\Psi}$, s.t.
\begin{itemize}
\item $\calb\defas\calg\cup\calr\cup\cala$ is a set of atomic
  propositions, where
$\calg\defas\{G_1,...,G_N\}$,
$\calr\defas\{R_1,...,R_K\}$,
$\cala\defas\{A_1,...,A_M\}$ are respectively sets of 
{goal},
{refinement} and
{domain-assumption labels}.
We denote with \cale the set of element labels:
$\cale\defas\calg\cup\cala$;
\item \caln is a set of numerical variables in the rationals;

\item \cald is an and-or directed acyclic graph  of
\emph{elements} in \cale 
(or nodes) and \emph{refinements} in \calr (and nodes);

\item $\Psi$ is a \smtlarat formula on \calb and \caln, representing 
the conjunction of all relation edges, user-defined constraints and
assertions. 

\end{itemize}

\ignoreinshort{
\noindent
The structure of a CGM is an and-or directed acyclic graph (DAG) of
\emph{elements}, as 
nodes, and \emph{refinements}, as (grouped) edges, which are labeled
by atomic propositions and can be augmented with arbitrary constraints
in form of graphical relations and Boolean or \smtlarat{} formulas
--typically conjunctions of smaller global 
and local constraints-- on the element and refinement labels and on
the numerical variables.  
}
\ignoreinshort{
Notice that each non-leaf element $E$ is implicitly or-decomposed into 
the set of its incoming refinements $\set{R_i}$ (i.e., $E \iff (\bigvee_i R_i)$)
and that each refinement $R$
is and-decomposed into the set of its source elements $\set{E_j}$ 
(i.e., $R \iff (\bigwedge_j E_j)$).~
Intuitively, a CGM describes a (possibly complex) combination of
alternative ways of realizing a set of requirements in terms 
of a set of tasks, under certain domain assumptions.
}
%

\ignoreinshort{Let $\calm\defas\tuple{\calb,\caln,\cald,\Psi}$ be a CGM.
A \new{realization} $\mu$ of \calm 
is an assignment of truth values to \calb and of rational
values to \caln (aka, a \larat-interpretation)
which:}
\ignoreinlong{\noi
A \new{realization} $\mu$ of \calm 
is an assignment of values to \calb and \caln s.t.:}

\begin{aenumerate}

\item for each non-leaf element $E$, $\mu$ satisfies $\bigl( E \iff
  (\bigvee_{R_i \in \rm{RefinementsOf}(E)} R_i)\bigr) $ --i.e., 
$E$ is part of a realization $\mu$ if and only if one of
  its refinements is in $\mu$;

\item 
for each refinement
$\refines{\bigl(\enum{E}{n}\bigr)}{E}$, 
$\mu$ satisfies $((\bigwedge_{i=1}^{n} E_i) \equivalent R)$ --i.e., $R$
is part of $\mu$ iff and only if all   of its sub-elements $E_i$  are in  $\mu$;


\item 
$\mu$ satisfies $\Psi$ --i.e., the elements and refinements occurring in $\mu$, and the values
  assigned by $\mu$ to the numerical attributes,  comply with 
all the relation edges, the 
user-defined constraints and user assertions  in $\Psi$. 
\end{aenumerate}
\noi
We say that an element $E$ or refinement $R$ is
\emph{satisfied} \resp{\emph{denied}} in $\mu$ if it is assigned 
to $\top$ \resp{$\bot$} by $\mu$.
$\mu$ 
is represented graphically as the  sub-graph of \calm where 
all the denied element and refinement nodes are eliminated. 
%
\ignoreinshort{We say that \calm, including user assertions, 
is \new{realizable} if it has at least one realization, 
 \new{unrealizable}
otherwise.}

As described in \cite{nguyensgm16}, a CGM \calm is encoded into a
\smtlarat formula $\Psi_{\calm}$, and the user preferences into
\new{numerical objective functions} \set{obj_1,...,obj_k}, which are
fed to the OMT solver \optimathsat, which returns optimal
solutions wrt. \set{obj_1,...,obj_k}, which are then converted back by
CGM-tool into optimal realizations.

%% file: evolutionreq_reasoning.tex

\fakesubsubsection{Evolution Requirements}
Here we formalize the notions described in \sref{sec:evolutionreq}.
Let $\calmone\defas \tuple{\calb_1,\caln_1,\cald_1,\Psi_1}$ be the
original model, $\mu_1$ be some  realization of
$\calmone$ and $\calmtwo\defas
\tuple{\calb_2,\caln_2,\cald_2,\Psi_2}$ be a new version of \calmone.
We look for a novel realization \mutwo for \calmtwo. 

\ignore{
We assume that the system of the original model $\calmone$
is already built based on the realization $\mu_1$.  Over the time, the
model $\calmone$ evolves and becomes a new model $\calmtwo\defas
\tuple{\calb_2,\caln_2,\cald_2,\Psi_2}$. In the evolution requirement
engineering problem, we want to find a realization $\mu_2$ for
$\calmtwo$ such that we can make the most use out of $\mu_1$. The
definition for "most use" can be varied, it can be the familiarity
between the two realization $\mu_1$ and $\mu_2$, or the effort needed
to implement the system from $\mu_2$ provided that we already have
$\mu_1$ implemented. }

Stakeholders can select a subset of the elements, called \new{elements
  of interest}, on which to focus, which can be 
requirements, tasks, domain assumptions, and intermediate goals. 
(When not specified otherwise, we will assume by default that all
elements are of interest.)
Let $\cale^* \subseteq \cale_1 \cup \cale_2$ be
the subset of the {elements of interest}, and let
$\cale^*_1 \defas \cale^* \cap \cale_1$ and $\cale^*_2 \defas \cale^*
\cap \cale_2$ be the respective subsets of $\calmone$ and
$\calmtwo$. We define $\cale^*_{common}  \defas  \{E_i \in \cale^*_2
\cap \cale^*_1\}$ as the set of elements of interest 
occurring in both \calmone and \calmtwo, and 
$\cale^*_{new}  \defas  \{E_i \in \cale^*_2 \setminus \cale^*_1\}$
as the set of new elements of interest in $\calmtwo$.

\paragraph{Familiarity.}
In its simplest form, the cost of familiarity can be defined as follows:
\begin{eqnarray}
%
\label{eq:addedcost_1}
 \familiaritycostof{\mu_2}{ \mu_1}& \defas & \mid \ \{E_i \in
 \cale^*_{common} \ \mid \ \mu_2(E_i) \neq \mu_1(E_i)\}  \ \mid 
\\
\label{eq:addedcost_2}
 & + &\mid \ \{E_i \in \cale^*_{new} \  \mid \  \mu_2(E_i)=\top\}\ \mid,
\end{eqnarray}
\noi where $\mid S\mid$ denotes the number of elements of a set $S$.
\noi
\familiaritycostof{\mu_2}{ \mu_1} is the sum of two components:
\begin{itemize}
\item[\eqref{eq:addedcost_1}] 
 the number of common elements of interest (e.g.,
tasks) which were in \muone and are no more in \mutwo, 
plus the number of these which were not in \muone and now are in
\mutwo, 
\item[\eqref{eq:addedcost_2}] 
the number of new elements of interest which are in \mutwo.
%
\end{itemize}
In a more sophisticate form, each element of interest $E_i$ can be given
some rational weight value $w_i$~\footnote{Like {\sf Penalty}, {\sf
  Cost} and {\sf WorkTime} in Figure~\ref{fig:CGMLex}.}, so that 
the cost of familiarity can be defined as follows:
\begin{eqnarray}
%
\label{eq:weightedaddedcost_1}
 \weightedfamiliaritycostof{\mu_2}{ \mu_1}& \defas & 
\sum_{E_i \in\cale^*_{common}} w_i \cdot {\sf Int}(\mu_2(E_i) \neq \mu_1(E_i))
\\
%
\label{eq:weighedtaddedcost_2}
 & + &\sum_{E_i \in\cale^*_{new}} w_i \cdot {\sf Int}(\mu_2(E_i)=\top),
\end{eqnarray}
\noi
where {\sf Int()} converts {\sf true} and {\sf false} into the
values $1$ and $0$ respectively.

Both forms are implemented in CGM-Tool. 
(Notice that \eqref{eq:addedcost_1} and \eqref{eq:addedcost_2}, 
or even \eqref{eq:weightedaddedcost_1} and \eqref{eq:weighedtaddedcost_2},  can
also be set as distinct objectives in CGM-Tool.)
Consequently, a realization \mutwo maximizing familiarity 
is produced by invoking the OMT solver on the formula
$\Psi_{\calmtwo}$ and the objective
\familiaritycostof{\mu_2}{ \mu_1} or
\weightedfamiliaritycostof{\mu_2}{\mu_1} to minimize. 

\paragraph{Change effort.}
We restrict the elements of interest to tasks only.
In its simplest form, the change effort can be defined as follows:
\begin{eqnarray}
\label{eq:addedcost_4}
 \effortcost{\mu_2}{\mu_1}& \defas & 
\mid \ \{T_i \in \cale^*_{common}\ \mid \ \mu_2(T_i) = \top, \text{
  and } \mu_1(T_i) = \bot\} \ \mid
\\
\label{eq:addedcost_5}
 &+&  \mid \ \{T_i \in
 \cale^*_{new} \ \mid \  \mu_2(T_i) = \top\} \ \mid. 
 \end{eqnarray}
\effortcost{\mu_2}{\mu_1} is the sum of two components:
\begin{itemize}
\item[\eqref{eq:addedcost_4}] is the number of common tasks which were not
in \muone and which are now in \mutwo,
\item[\eqref{eq:addedcost_5}]  is the number of  new tasks which
are in \mutwo.
\end{itemize}

As above, in a more sophisticate form, each task of interest $T_i$
can be given 
some rational weight value $w_i$, so that 
the change effort can be defined as follows:
\begin{eqnarray}
%
\nonumber
\label{eq:weightedaddedcost_4}
 \weightedeffortcostof{\mu_2}{ \mu_1}& \defas & 
\sum_{T_i \in\cale^*_{common}} w_i \cdot {\sf Int}(\mu_2(T_i)=\top)
\cdot {\sf Int}(\mu_1(T_i)=\bot)
\\
\nonumber
\label{eq:weighedtaddedcost_5}
 & + &\sum_{T_i \in\cale^*_{new}} w_i \cdot {\sf Int}(\mu_2(T_i)=\top).
\end{eqnarray}
Both forms are implemented in CGM-Tool. 
Consequently, a novel realization \mutwo minimizing change effort
is produced by invoking the OMT solver on the formula
$\Psi_{\calmtwo}$ and the objective
\effortcost{\mu_2}{ \mu_1}
 or
\weightedeffortcostof{\mu_2}{ \mu_1}.

Notice an important difference between \eqref{eq:addedcost_1} and
\eqref{eq:addedcost_4}, even if the former is restricted to tasks
only: a task which is satisfied in $\muone$ and is no more in $\mutwo$
worsens the familiarity of \mutwo wrt. \muone \eqref{eq:addedcost_1},
but it does not affect its change effort \eqref{eq:addedcost_4},
because
it does not require implementing one more task.


\fakesubsubsection{Comparison wrt. previous approaches}
Importantly, Ernst et al. \cite{ErnstBMJ12} proposed two similar 
notion of familiarity 
and change effort for (un-)constrained goal graphs:
\begin{description}
\item{\em familiarity}: maximize (the cardinality of) 
the set of tasks used in the previous solution;
\item{\em change effort}: (i) minimize (the cardinality of) 
the set of new tasks in the novel realization --or, alternatively, (ii)
minimize also the number of tasks. 
\end{description}
We notice remarkable differences of our approach wrt. the one in
\cite{ErnstBMJ12}. 

First, our notion of familiarity presents the following novelties:
\begin{renumerate}
\item 
it uses all kinds of elements, on stakeholders' demand, 
rather than only tasks;
\item 
it is (optionally) enriched also with \eqref{eq:addedcost_2};
\item  
\eqref{eq:addedcost_1} is sensitive also to tasks 
which were in the previous realization and which are not in the novel
one, since we believe that also these elements affect familiarity. 
\end{renumerate}
Also, in our approach both familiarity and change effort allow 
for adding \new{weights} to tasks/elements, and to combine  
familiarity and change-effort objectives lexicographically with other
user-defined objectives.

Second, unlike with \cite{ErnstBMJ12}, in which 
the optimization procedure is hardwired, we rely on logical encodings
of novel objectives into \omtlarat objectives, using \optimathsat as
workhorse reasoning engine. Therefore, new objectives 
require implementing no new reasoning procedure, only new \omtlarat encodings. 
For instance, 
we could easily implement also the notion of familiarity 
of \cite{ErnstBMJ12} by asking \optimathsat to minimize the objective:
\mbox{$\mid  \{T_i \in \cale^*_{common}\ \mid \ \mu_2(T_i) = \bot, \text{
  and } \mu_1(T_i) = \top\}  \mid$.}

Third, our approach deals with CGMs, 
which are very expressive formalisms,  are enriched by
Boolean and numerical constraints, and are supported by a tool
(CGM-Tool) with efficient search functionalities for
optimum realizations. These functionalities, which are 
enabled by state-of-the-art SMT and OMT technologies
\cite{sebastiani15_optimathsat,st_cav15},  
 scale very well, up to thousands of
elements, as shown in the empirical evaluation of
\cite{nguyensgm16}. In this paper we further enrich these 
functionalities so that to deal also with evolving CGMs and evolution
requirements.  

Fourth, unlike with \cite{ErnstBMJ12}, where 
realizations are intrinsically supposed to be {\em minimal},
 in our approach minimality is an objective stakeholders can set and obtain
as a byproduct of {\em minimum} solutions, but it is not mandatory. 
This fact is relevant when dealing with familiarity evolution
requirements, because objective \eqref{eq:addedcost_1} can
conflict with minimality, because it may force
the presence of tasks from the previous solution which have become
redundant in the new model. Thus, sometimes CGM-tool may return a non-minimal
model if the stakeholder prioritizes familiarity above all other
objectives. 



\FUTUREWORK{
\RSTODO{Say something about ``structurally minimal'' CGMs?\\
impose $(G\imp {\sf oneof(R_1,...,R_k)})$ for each non-leaf goal
and $(G\imp {\sf \bigvee_{R_i \in {\sf fathersOf(G)}}R_i)}$ }}

%% file: implementation.tex
\JMSUGGESTION{Section 4 would run the different req problems and
produce solutions. We can report time required and perhaps do some
senstivity analysis, e.g. if we make small changes to costs etc, do we get a
drastically different solution or a very similar one.}

\ignore{
\begin{figure}[t]
\centering 
\includegraphics[height=0.39\textheight, width=\textwidth]{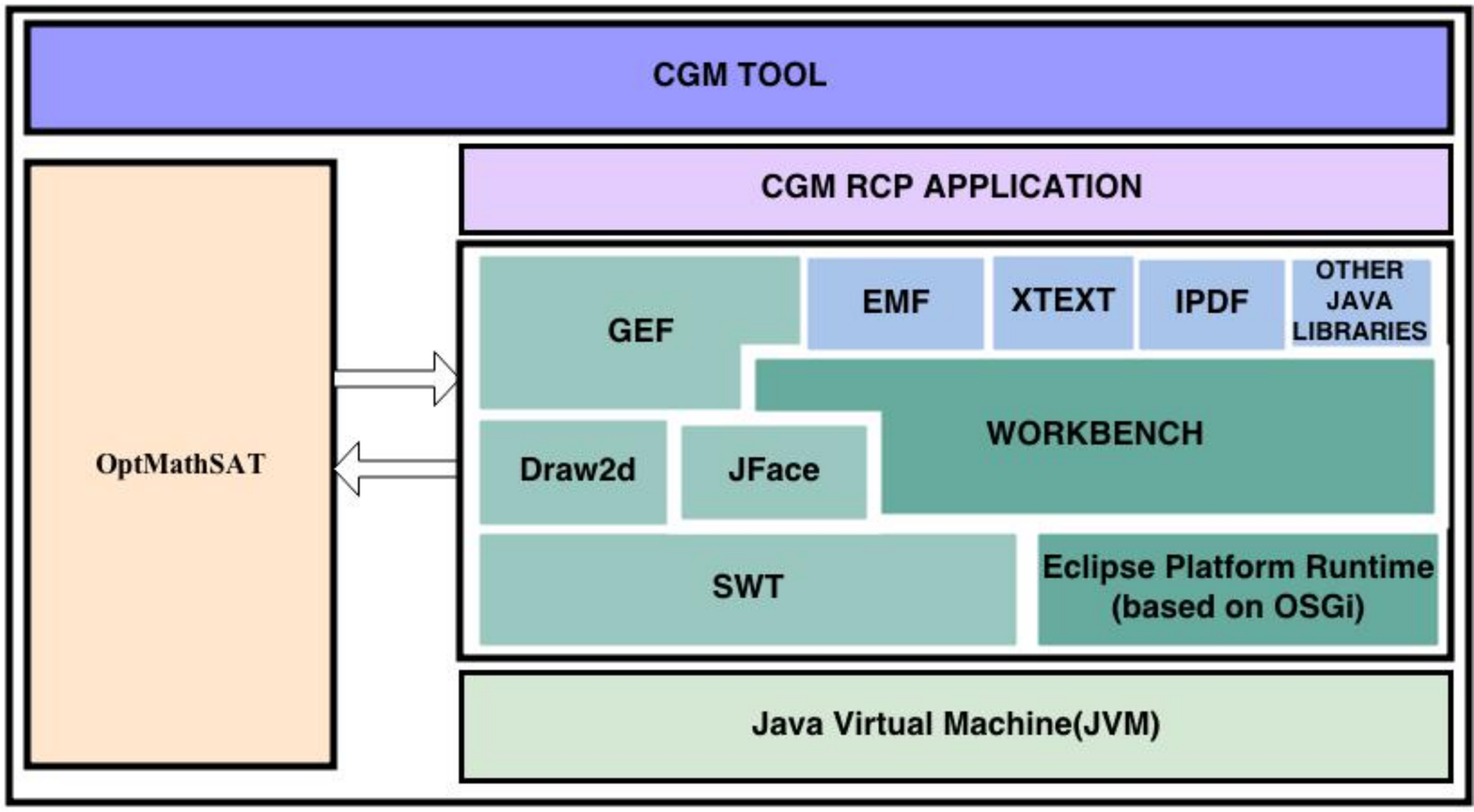}
\caption{\label{figCGM--Tool} CGM-Tool Component view}
\end{figure}
}

CGM-Tool provides support for modeling and reasoning on
CGMs \cite{nguyensgm16}.  Technically, CGM-Tool is a standalone
application written in Java and its core is based on Eclipse RCP
engine. Under the hood, it encodes CGMs and invokes
the OptiMathSAT%
~\footnote{\url{http://optimathsat.disi.unitn.it}} OMT solver
 \cite{st_cav15}\ignoreinshort{ to support reasoning on CGMs}. It is freely
 distributed \ignoreinshort{as a compressed archive file} for multiple platforms
 \footnote{\url{http://www.cgm-tool.eu/}}. \ignoreinshort{Currently CGM-Tool supports
 the functionalities in \cite{nguyensgm16}: 
\begin{description}
	\item[Specification of projects:] CGMs
are created within the scope of project containers. A project contains
a set of CGMs that can be used to generate  reasoning sessions with
OptiMathSAT (i.e., scenarios);	
\item[Diagrammatic modeling:] the tool enables the creation of CGMs as
  diagrams; it provides real-time check for refinement cycles and reports invalid  links;
\item[Consistency/well-formedness check:] CGM-Tool provides
 the ability to run consistency analysis and well-formedness checks on the
  CGMs;	
\item[Automated Reasoning:] CGM-Tool provides the automated reasoning
  functionalities {mentioned in section \ref{sec:background_ex}}, and
  described in detail in \cite{nguyensgm16}.
\end{description}
}

\ignoreinshort{
\noi
With this work, we have enhanced CGM-Tool with
the following functionalities:
 \begin{description}
\item[Evolution Requirements Modelling and Automated Reasoning:] by means of scenarios,
  stakeholders can generate \new{evolution sessions}, which allows for (i) defining
  the first model and finding the first optimal
  realization, (ii) modifying the model to
  obtain the new models, and (iii) generating automatically 
  the ``similar'' realization (as discussed in
  section \ref{sec:evolutionreq}).  
\end{description}
}
\ignoreinlong{
With this work, we have enhanced CGM-Tool with
the functionalities for 
\new{Evolution Requirements Modelling and Automated
  Reasoning:} 
  stakeholders can generate \new{evolution sessions}, which allows for (i) defining
  the first model and finding the first optimal
  realization, (ii) modifying the model to
  obtain the new models, and (iii) generating automatically 
  the ``similar'' realization (as discussed in
  section \ref{sec:evolutionreq}).  
}

As a proof of concept, we have performed various attempts on variants 
of the CGM of \sref{sec:evolving}.
The automated generation of the realizations always required
negligible amounts of CPU time, like those reported in \sref{sec:evolutionreq}.

\ignore{
CGM-Tool extends the STS-Tool~\cite{paja:2012:ststool:re} as an RCP application by using the major frameworks shown in Figure~\ref{figCGM--Tool}: {\em Rich Client Platform (RCP)}, a platform for building rich client applications, made up of a collection of low level frameworks such as OSGi, SWT, JFace and Equnix, which provide us a workbench where to get things  like menus, editors and views; 
{\em Graphical Editing Framework (GEF)}, a framework used to create
graphical editors for graphical modeling tools (e.g., tool palette and
figures which can be used to graphically represent the underlying data
model concepts); {\em Eclipse Modeling Framework (EMF)}, a modeling
framework and a code generation facility for building tools and
applications based on a structured data model. 
}


%% file: related.tex
The most relevant work to our proposal can be found in Neil Ernst's PhD thesis \cite{Ernst1}, also published in \cite{ErnstBJ11,ErnstBMJ12}. In this work, Ernst proposes to solve the requirements evolution problem by searching for specifications (i.e., solutions) for the evolved requirements that satisfy some desire property relative to the old solution. His proposal for possible desired properties include minimal change effort, maximal familiarity, and solution reuse over the history of changes. Ernst also proposes to use Techne \cite{JuretaBEM10} to provide a precise formal specification of a minimal requirement engineering knowledge base (REKB) which then can be used by another problem solver as a tool for storing information acquired during requirements acquisition and domain modelling, as well as justifying problem decomposition; and asking a variety of question that can help compute and compare alternative solutions. The major difference between his work and our proposal is that the language he uses to model requirements (essentially, Propositional Logic) is not expressive enough to capture evolution requirements, so he needed to implement algorithms that find specifications for a given requirements model, and search among those to find ones that satisfy evolution requirements (minimal effort, maximal familiarity, etc.)

Evolving requirements models in order to handle unanticipated changes can be considered requirements management. Here, evolution is treated as the addition/deletion of requirements leading to a new requirements problem, and the re-calculation of a new solution, with no reference to the old one. A common approach to requirements management is impact analysis, which can be used to workbench different scenarios, as done in the AGORA tool \cite{Kaiya02}.

Working with existing requirements has been studied in the area of software product lines and feature models. Evolving a single set of requirements over time shares many similarities with the problem of maintaining several sets of requirements for different products in a product family. There has been a number of papers looking at the problem of automated reasoning with feature models \cite{Schobbens}. Most of this work conducts consistency checking, to determine whether a given configuration of features is satisfiable, instead of enumerating all solutions and looking for optimal ones. Tun et al. [Tun] use problem frames to incrementally model sets of features for a product line. Temporal logic is used to minimize feature interaction. The chief difference between product line approaches and our work is that we add evolution requirements and look for solution that fulfill new requirements and are optimal relative to the desired evolution property (minimal effort etc.)

%% file: concl.tex
We have proposed to model changing requirements in terms of changes to CGMs. Moreover, we have introduced a new class of requirements (evolution requirements) that impose constraints on allowable evolutions, such as minimizing (implementation) effort or maximizing (user) familiarity. We have demonstrated how to model such requirements in terms of CGMs and how to reason with them in order to find optimal evolutions. 

Our future plans for this work include further evaluation with larger case studies, as well as further exploration for new kinds of evolution requirements that can guide software evolution.

%% file: main.bbl
\begin{thebibliography}{10}

\bibitem{barrettsst09}
C.~W. Barrett, R.~Sebastiani, S.~A. Seshia, and C.~Tinelli.
\newblock {Satisfiability Modulo Theories}.
\newblock In {\em Handbook of Satisfiability}, chapter~26, pages 825--885. IOS
  Press, 2009.

\bibitem{Ernst1}
N.~A. Ernst.
\newblock {\em Software Evolution: A Requirements Engineering Approach}.
\newblock PhD thesis, University of Toronto, 2012.

\bibitem{ErnstBJ11}
N.~A. Ernst, A.~Borgida, and I.~Jureta.
\newblock Finding incremental solutions for evolving requirements.
\newblock In {\em RE}, pages 15--24. IEEE, 2011.

\bibitem{ErnstBMJ12}
N.~A. Ernst, A.~Borgida, J.~Mylopoulos, and I.~Jureta.
\newblock {Agile Requirements Evolution via Paraconsistent Reasoning}.
\newblock In J.~Ralyt{\'e}, X.~Franch, S.~Brinkkemper, and S.~Wrycza, editors,
  {\em CAiSE}, volume 7328 of {\em Lecture Notes in Computer Science}, pages
  382--397. Springer, 2012.

\bibitem{JuretaBEM10}
I.~Jureta, A.~Borgida, N.~A. Ernst, and J.~Mylopoulos.
\newblock Techne: Towards a new generation of requirements modeling languages
  with goals, preferences, and inconsistency handling.
\newblock In {\em RE}, pages 115--124. IEEE Computer Society, 2010.

\bibitem{Kaiya02}
H.~Kaiya, H.~Horai, and M.~Saeki.
\newblock Agora: Attributed goal-oriented requirements analysis method.
\newblock {\em 2014 IEEE 22nd International Requirements Engineering Conference
  (RE)}, 0:13, 2002.

\bibitem{Lehman80}
M.~M. Lehman.
\newblock {Programs, Life Cycles, and Laws of Software Evolution}.
\newblock In {\em Proceedings of the IEEE}, pages 1060--1076, Sept. 1980.

\bibitem{nguyensgm16}
C.~M. Nguyen, R.~Sebastiani, P.~Giorgini, and J.~Mylopoulos.
\newblock Multi object reasoning with constrained goal model.
\newblock {\em CoRR}, abs/1601.07409, 2016.
\newblock Under journal submission. Available as
  \url{http://arxiv.org/abs/1601.07409}.

\bibitem{Schobbens}
P.-Y. Schobbens, P.~Heymans, J.-C. Trigaux, and Y.~Bontemps.
\newblock Generic semantics of feature diagrams.
\newblock {\em Comput. Netw.}, 51(2):456--479, Feb. 2007.

\bibitem{sebastiani15_optimathsat}
R.~Sebastiani and S.~Tomasi.
\newblock {Optimization Modulo Theories with Linear Rational Costs}.
\newblock {\em ACM Transactions on Computational Logics}, 16(2), March 2015.

\bibitem{st_cav15}
R.~Sebastiani and P.~Trentin.
\newblock {OptiMathSAT: A Tool for Optimization Modulo Theories.}
\newblock In {\em Computer-Aided Verification, CAV}, volume 9206 of {\em LNCS}.
  Springer, 2015.

\bibitem{souza:phdthesis12}
V.~E.~S. Souza.
\newblock {\em {Requirements-based Software System Adaptation}}.
\newblock Phd thesis, University of Trento, 2012.

\end{thebibliography}
